\documentclass[twoside,preprint2]{aastex}
\usepackage[english]{babel}
\usepackage{graphics}
\usepackage{psfig}
\usepackage{latexsym,amssymb}

\newcommand{\teff}{$T_{\mathrm{eff}}$}
\newcommand{\mteff}{T_{\mathrm{eff}}}
\newcommand{\logg}{$\log{g}$}
\newcommand{\mlogg}{\log{g}}
\newcommand{\metal}{[M/H]}
\newcommand{\mmetal}{\mathrm{[M/H]}}
\newcommand{\etal}{{\it et al.\ }}

\shortauthors{E. Bertone et al.}
\shorttitle{A combined analysis of ATLAS and NextGen model atmospheres}

\begin{document}

\title{ATLAS vs.\ NextGen model atmospheres:\\
a combined analysis of synthetic spectral energy distributions}

\author{E. Bertone}

\affil{Instituto Nacional de Astrof\'\i sica \'Optica y Electr\'onica, Luis
Enrique Erro 1, 72840 Tonantzintla, Mexico}

\author{A. Buzzoni}

\affil{INAF - Osservatorio Astronomico di Bologna, Via Ranzani 1 40127 Bologna, Italy}

\author{M. Chavez\altaffilmark{1}, L. H. Rodriguez-Merino}

\affil{Instituto Nacional de Astrof\'\i sica \'Optica y Electr\'onica, Luis
Enrique Erro 1, 72840 Tonantzintla, Mexico}

\altaffiltext{1}{Visiting Astronomer at the Steward Observatory and Lunar and
Planetary Laboratory, University of Arizona, Tucson, AZ85721.}

\begin{abstract}
We carried out a critical appraisal of the two theoretical models, Kurucz'
ATLAS9 and PHOENIX/NextGen, for stellar atmosphere synthesis. Our tests relied
on the theoretical fit of spectral energy distributions (SED) for a sample of 334 target stars along the whole spectral-type
sequence, from the classical optical catalogs of \citet{gs83} and
\citet{jhc84}. The best-fitting physical parameters (\teff, \logg) of stars
allowed an independent calibration of the temperature and bolometric scale
vs.\ empirical classification parameters (i.e.\ spectral type and MK
luminosity class); in addition, the comparison of the synthetic templates from
the ATLAS and NextGen grids allowed us to probe the capability of the models
to match spectrophotometric properties of real stars and assess the impact of
the different input physics. We can sketch the following main conclusions of
our analysis:

{\it i)} fitting accuracy of both theoretical libraries drastically degrades
at low \teff, where both ATLAS and NextGen models still
fail to properly account for the contribution of molecular features in the
observed SED of K-M stars.

{\it ii)} Comparing with empirical calibrations, both ATLAS and NextGen fits
tend, in average, to predict slightly warmer (by 4--8\%) \teff\ for both giant
and dwarf stars of fixed spectral type, but ATLAS provides in general a
sensibly better fit (a factor of two lower $\sigma$ of flux residuals) than
NextGen.

{\it iii)} There is a striking tendency of NextGen to label target stars with 
an effective temperature and surface gravity in excess with respect to ATLAS. 
The effect is especially evident for MK I-III objects, where a fraction of
stars of about one in four is clearly misclassified by NextGen in \logg. This
is a consequence of some ``degeneracy'' in the solution
space, partly induced by the different input physics and geometry constraints
in the computation of the integrated emerging flux (ATLAS model atmospheres
assume standard plane-parallel layers, while NextGen adopts, for low-gravity
stars, a spherical-shell geometry). 
A different $T(\tau)$ vertical structure of stellar atmosphere
seems also required for NextGen synthetic SEDs in order to better account for
limb-darkening effects in cool stars, as supported by the recent observations
of the EROS BLG2000-5 microlensing event.
\end{abstract}


\maketitle


\section{Introduction}

Theoretical computation of model atmospheres has been a leading issue of
stellar astrophysics in the last decades. In this framework, Kurucz'
\citeyearpar{kurucz70,kurucz79} pioneering work certainly stands as a main
reference, together with a few other major contributions like those of
\citet{gustafssonetal75} and \citet{tsuji76}, on the synthesis of red giant
stars.

In its more recent versions, Kurucz' \citeyearpar{kurucz92a,kurucz95} ATLAS
code included over 58 million spectral lines, providing an accurate
description of blanketing effects, that modulate ultraviolet emission of stars
\citep{holweger70, gustafssonetal75}, and also including nearly all the most
important di-atomic molecules, that shape spectral energy distribution (SED)
at longer wavelength. The lack of tri-atomic molecules ({\it in primis}
H$_2$O), and an incomplete treatment of TiO opacity, however, still prevents a
satisfactory match of stars cooler than 3500~K
\citep{kurucz92b,castellietal97}. This limit of ATLAS theoretical
atmospheres unfortunately affects a number of physical applications dealing,
for instance, with the study of cool pulsating variables, or the match to
integrated SED of galaxies, through stellar population synthesis.

More recently, \citet{hau99a,hau99b} have presented their PHOENIX/NextGen grid
of model atmospheres for dwarf and giant stars. \citet{allardetal00} extended
the original bulk of models to the pre-main-sequence (pre-MS) evolution at
the low-mass regime. With its 500 million atomic and molecular lines and a
spherical geometry treatment of stellar structure, the NextGen library is 
arguably the most advanced one currently available in the literature. As the
low-temperature physical regime is suitably sampled, with models as cool as
$\mteff = 2000$~K, these may possibly fill the gap assuring a homogenous
coverage of the stellar fundamental parameters across the whole H-R diagram.

Given the relevance of the Kurucz and Hauschildt \etal contributions, it could 
be of special interest, at this stage, to carry out a combined analysis of the
ATLAS vs.\ NextGen codes, in order to check their mutual capabilities to match
spectrophotometric properties of real stars and assess self-consistency in
their input physics. Our analysis follows the Hauschildt \etal  
\citeyearpar{hau99a} preliminary discussion, and will be carried out in two
steps. After a brief description of the main features of each theoretical
dataset (Sec.\ 2), we will first try a fit to observations of template stars
and compare model outputs (Sec.\ 3). This will rely on an original
optimization procedure, that also provides an estimate of the fit uncertainty
across the (\teff, \logg, \metal) phase space. The best fits will allow to
establish an effective temperature scale and a calibration of the bolometric
correction scale (Sec.\ 4). In Sec.\ 5 we will then analyse ATLAS vs.\ NextGen
theoretical models for fixed fiducial spectral types. A full
summary of the relevant conclusions of our tests is finally given in Sec.\ 6.


\section{Input model atmospheres and grid properties}
\label{sec:modatm}

ATLAS model atmospheres assume steady-state plane-parallel layers under the
hypothesis of local thermodynamic equilibrium (LTE). Line blanketing is
computed statistically by means of opacity distribution functions (ODF), that
average the contribution of the different atomic/molecular species through the
corresponding oscillator forces \citep{stromkurucz66,kurucz70,kurucz79}. For
our work, we used the ATLAS\,9 version \citep{kurucz95}, whose treatment of
convection is based on the mixing length theory \citep{bohmvitense58}
and accounts for the so-called ``approximate overshooting'', according to
\citet{castellietal97}. The mixing length parameter is set to $\ell/H_p=1.25$
and the microturbulence velocity to 2 km s$^{-1}$ throughout. The whole
theoretical library is made available at the Kurucz web
site.\footnote{$\;$\texttt{http://kurucz.harvard.edu}}

The model library spans a temperature range between $3500 \leq \mteff \leq
50\,000$~K, sampled at a variable step from 250 to 2500~K with increasing
temperature; surface gravity and metallicity cover the interval $0.0 \leq
\mlogg \leq 5.0$~dex at steps of $\Delta \mlogg = 0.5$ dex, and $-5.0 \leq
\mmetal \leq +1.0$~dex, respectively. The corresponding SEDs, which also account 
for line opacity through the ODFs, span from the
far-ultraviolet ($\lambda = 90$~\AA) to the far-infrared (160~$\mu$m), sampled
by 1221 wavelength points, with $\Delta\lambda = 10$~\AA\ in the UV and
$20$~\AA\ in the visual range.

NextGen models have originally been computed by Hauschildt and collaborators 
with the multipurpose code PHOENIX \citep[e.g.,][]{hau96}. They assume LTE and 
plane-parallel geometry for dwarf stars, while a spherical symmetry is adopted 
in low-gravity  model atmospheres ($\mlogg \leq 3.5$) for giant and pre-MS
stars \citep{hau99b, allardetal00}. As a striking difference with
respect to the Kurucz models, direct opacity sampling is performed including
over 500 million lines of atomic and molecular species along the spectrum.

The phase-space domain of the Next\-Gen grid spans the $2000 \leq \mteff \leq
10\,000$~K range at steps $\Delta\mteff = 100$--$200$~K, with gravity in the
interval $0.0 \leq \mlogg \leq 5.5$ ($\Delta \mlogg = 0.5$~dex), and
metallicity in the range $-4.0 \leq \mmetal \leq +0.3$. The SEDs cover the
wavelength range from 100~\AA\ to 970 $\mu$m sampled at coarser wavelength
steps, that can change from model to model (depending on the intervening
absorption features in the spectrum). However, a typical
$\Delta\lambda=2$~\AA\ step in the optical region can be picked out. All
the data are available via anonymous ftp.\footnote{$\;$\texttt{ftp://calvin.physast.uga.edu/pub/} \\
\texttt{http://dilbert.physast.uga.edu/$\sim$yeti} \\
Note that the libraries of dwarf and giant stars available at these sites
have lower \teff\ limits than the published ones.}

In addition to the standard fundamental parameters (i.e.\ \teff, \logg,
\metal), spherical models in NextGen require one supplementary ``dimension''
in the phase space. As $g \propto M/R^2$ and  $L \propto R^2T^4$, then the
emerging luminosity becomes $L \propto (M\ T^4)/g$. Contrary to the plane-parallel
case, therefore, mean surface brightness depends on the absolute size of stars 
through the stellar mass ($M$).
Giant-star models in the NextGen grid are computed for $5\ M_\odot$, while
pre-MS stars assume $0.1\ M_\odot$. However, mass scaling is found to induce 
second-order effects on the stellar SED \citep{hau99b}, although total
luminosity of the models scales of course as $L \propto M$ for
fixed temperature and gravity.


\section{Matching SED of template stars}
\label{sec:method}

A first ``sanity'' check in our analysis concerns the match with template
stars along the whole O~$\to$~M spectral-type sequence. Comparison with
observed SEDs is one of the most natural application of model
atmospheres. High-resolution spectra help in fact investigate chemical
composition of stars, while the shape of (pseudo-)continuum at lower
resolution gives clues on stellar gravity and effective temperature.

\subsection{The empirical spectral libraries}

For our test we considered the complete libraries of stellar spectra by
\citet[][hereafter GS83]{gs83} and \citet[][JHC84]{jhc84}. Both sets of
spectra span the whole range of stellar parameters, from giants (MK class
I--III) to dwarfs (MK class IV--V), and have been widely used in the
literature, particularly for population synthesis studies
\citep[e.g.,][]{pickles85,guiderd87,fanellietal87,bruzualcharlot93}. The
complete sample of target stars amounts to 336 objects (175 stars from GS83
and 161 from JHC84, with no stars in common to the two libraries).

The GS83 data cover a wide wavelength range, from 3130 to 10800~\AA, observed
at low resolution (FWHM=20~\AA\ in the blue and 40~\AA\ in the red) and
sampled at steps of 10-20~\AA. Due to a poorer S/N quality in the ultraviolet,
particularly for cool stars \citep{gs83}, only the wavelength interval for
$\lambda > 3500$~\AA\ is suitable for our analysis. Thanks to a better
sampling ($\Delta \lambda = 1.4$~\AA), the FWHM~$\sim 4.5$~\AA\ resolution of
JHC84 spectra is better exploited, giving a more detailed picture of the main
absorption features of template stars in the $3510 \leq \lambda \leq
7427$~\AA\ spectral range. For both the GS83 and JHC84 datasets we had to
reject several regions (namely, around 6840--7000~\AA, 7140--7350~\AA,
7560--7720~\AA, 8110--8360~\AA, and 8900--9800~\AA) affected by telluric bands
of O$_2$ and H$_2$O.

In our work we adopted the original MK spectral classification by GS83 and
JHC84. For 24 out of 26 unclassified objects we relied on the SIMBAD
database. All the spectra have been corrected for Galaxy reddening and
atmospheric extinction as reported in the original data sources.

\subsection {The theoretical spectral libraries}

The subsample of [M/H] = 0 model atmospheres has been used to match the
observations. This choice is consistent with the mean metallicity of the GS83 
and JHC84 stars. A systematic search from high-resolution abundance studies in
the literature actually provided a mean value of $[Fe/H] = -0.10 \pm 0.24$ for
67 stars in the GS83 sample, and $[Fe/H] = -0.06 \pm 0.19$ for 25 JHC84
stars.\footnote{The bulk of  metallicity estimates for our star sample comes
from the catalogs of \citet{cayreletal97,cayreletal01}; other references are
\citet{hy82,lu82,kj84,faberetal85,lg86,n86,bc88,lcl88,kn89,lb89,e91,e98,gcm91,t91,t99,x91,fj93,tff93,w94,bl96,bl97,cpo96,zs96,fc97,fm97,sbs97,tt98,a99,bel99,cc01,gg01,ha01,ak02,v02},
based on different spectroscopic or photometric methods. Note that the only
two metal-poor stars (HD 94028 and SAO 102986) in the JHC84 catalog have been
excluded in our analysis so that we are eventually left with 334 objects in
the GS83 plus JHC84 total sample.}

\begin{figure}[!t]
\psfig{file=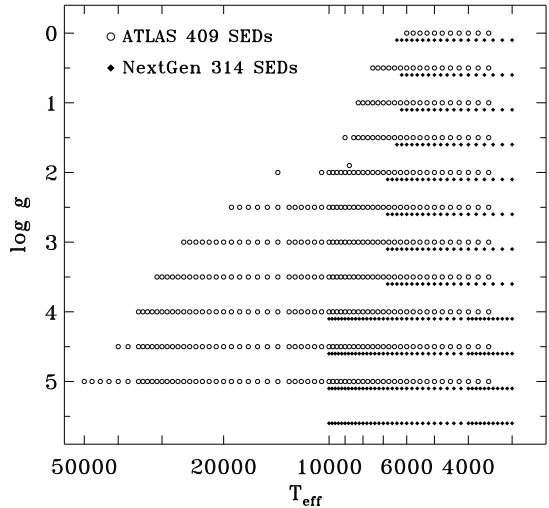,width=\hsize}
\caption{The grid of 409 ATLAS and 314 NextGen theoretical models considered
in our analysis. A solar metallicity has been assumed throughout.}
\label{fig:grids}
\end{figure}

A total of 409 theoretical SEDs have been collected from the ATLAS grid, while
from the NextGen library we extracted a set of 314 theoretical flux
distributions composed by 164 plane-parallel models for dwarfs with $3000 \le
T_\mathrm{eff} \le 10\,000$~K and $\log{g} > 3.5$ plus 150 spherical models
for giants with $3000 \le T_\mathrm{eff} \le 6800$~K and $\log{g} \le
3.5$. The (\teff, \logg)-space coverage of the two grids is shown in
Fig.~\ref{fig:grids}.

In order to consistently compare empirical and theoretical libraries, we
degraded both GS83 and JHC84 spectra with a Gaussian kernel of FWHM = 25~\AA,
rebinning the output at constant steps of $\Delta\lambda=5$~\AA. The same
procedure has been applied to the ATLAS and NextGen SEDs, sampled at the same
set of wavelength points. The effect of this
low-resolution approach on the results of our analysis (especially on the
calibration of the temperature scale), will be discussed in some detail in
Sec.~\ref{sec:teffbc} and \ref{sec:results}.

\subsection {Fitting procedure}
\label{sec:fittingproc}

A ``best fit'' for the (\teff, \logg) fundamental parameters (assuming $\mmetal
= 0$) was searched for each star in the GS83 and JHC84 samples  by matching the
observed SED with both ATLAS and NextGen libraries. As described in full
detail in \citet{bertone01} and \citet{bb01}, our method basically relies on a
minimization of the statistical variance in the relative flux domain, as a
measure of the similarity between target spectrum and theoretical SEDs across
the reference grid.

\begin{figure*}[!ht]
\begin{center}
\begin{tabular}{cc}
\includegraphics[width=7cm]{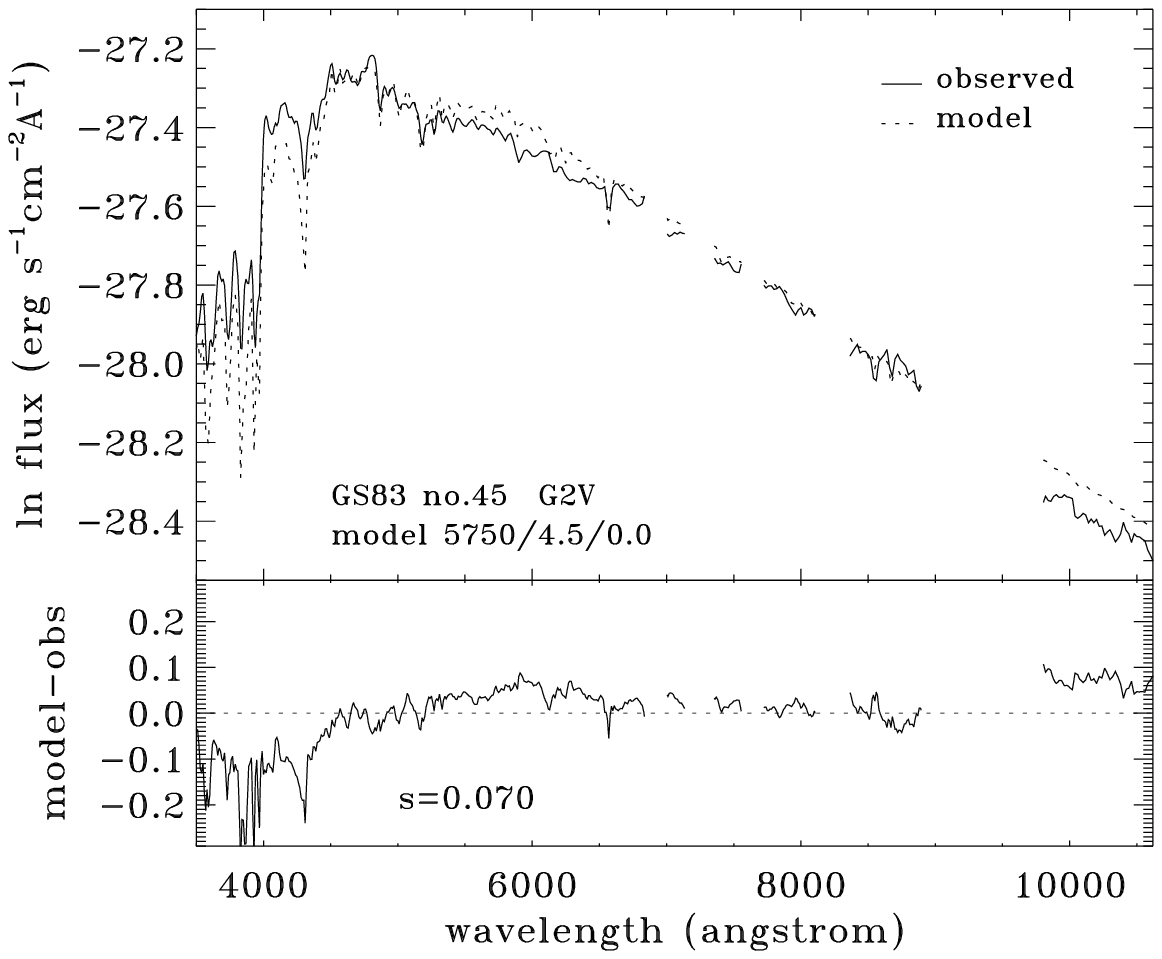} &
\includegraphics[width=7cm]{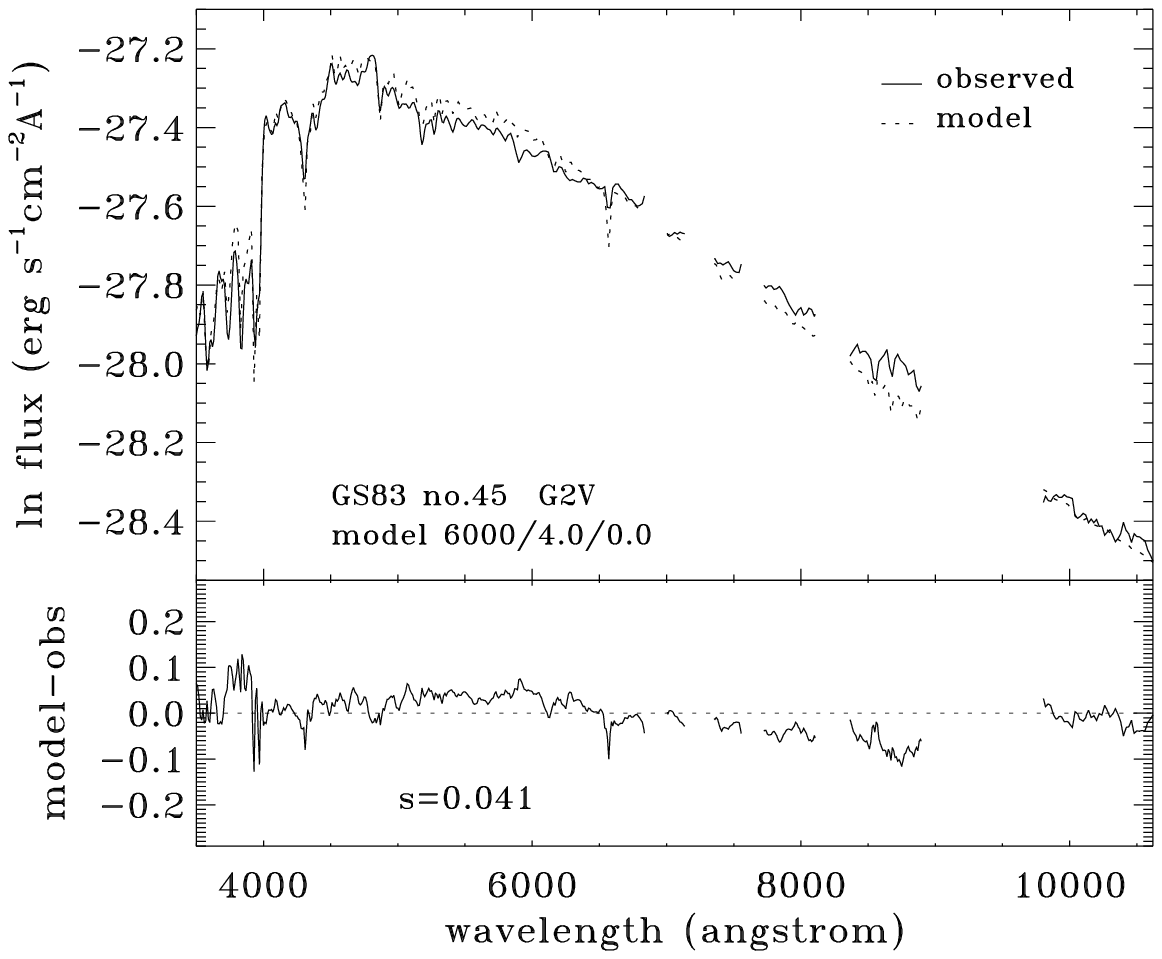} \\
\multicolumn{2}{c}{\includegraphics[width=7cm]{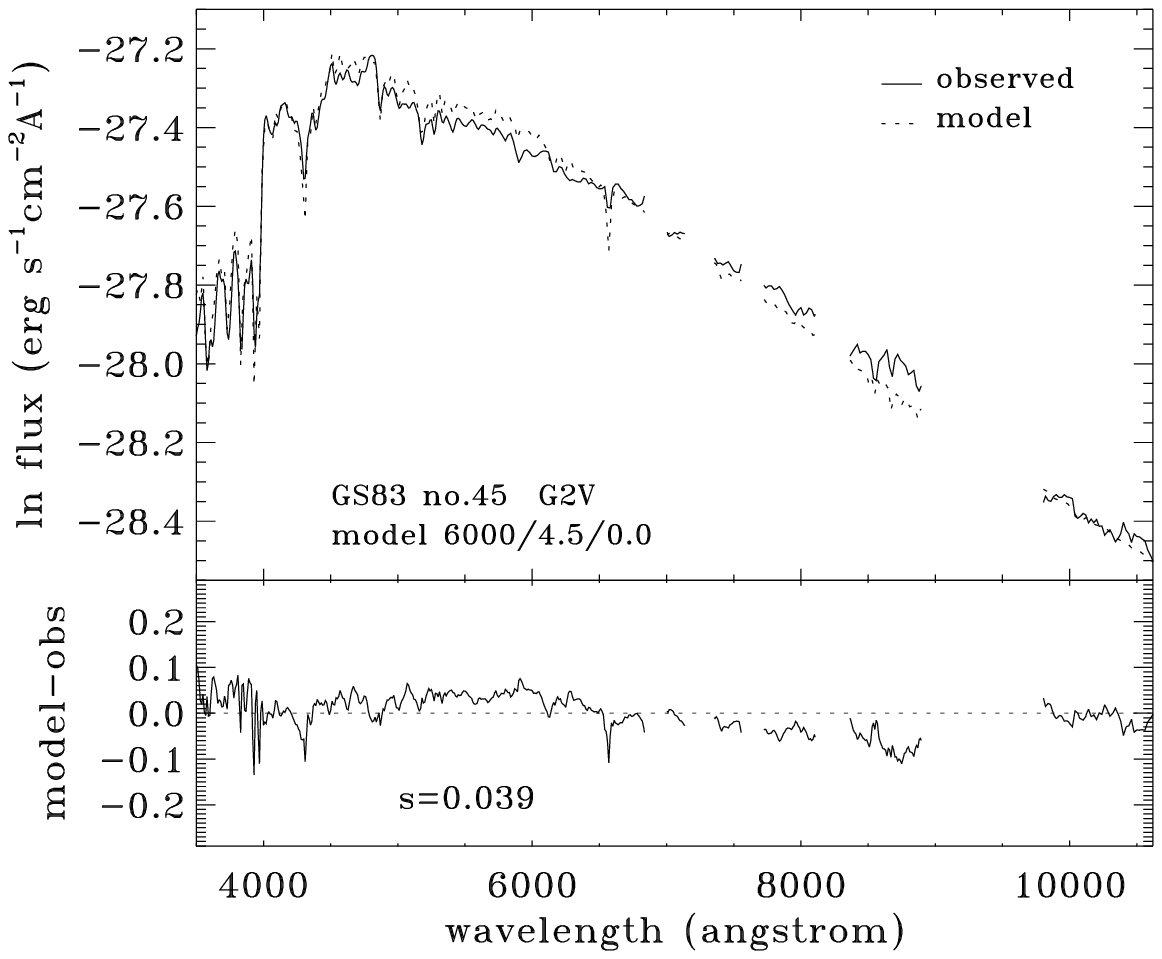}} \\ 
\includegraphics[width=7cm]{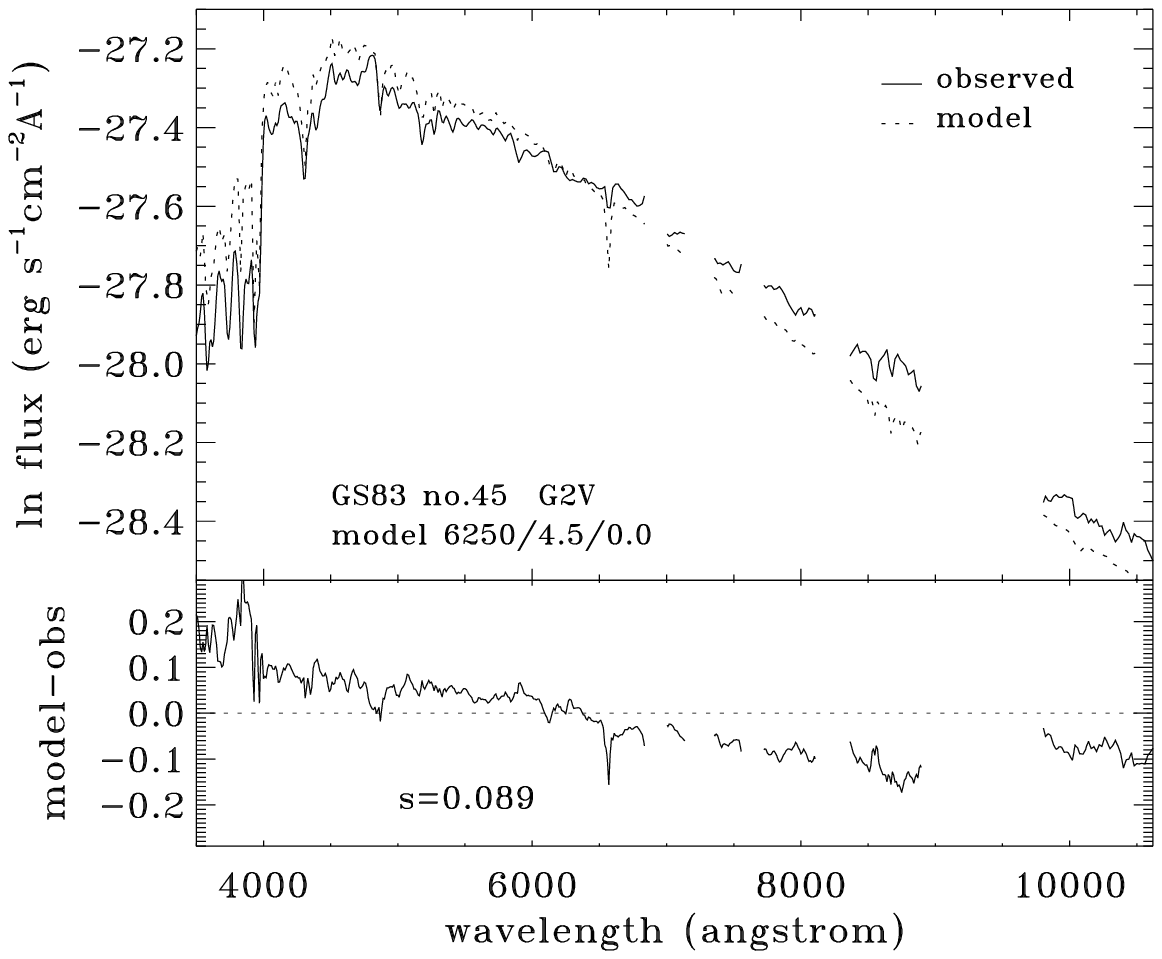} &
\includegraphics[width=7cm]{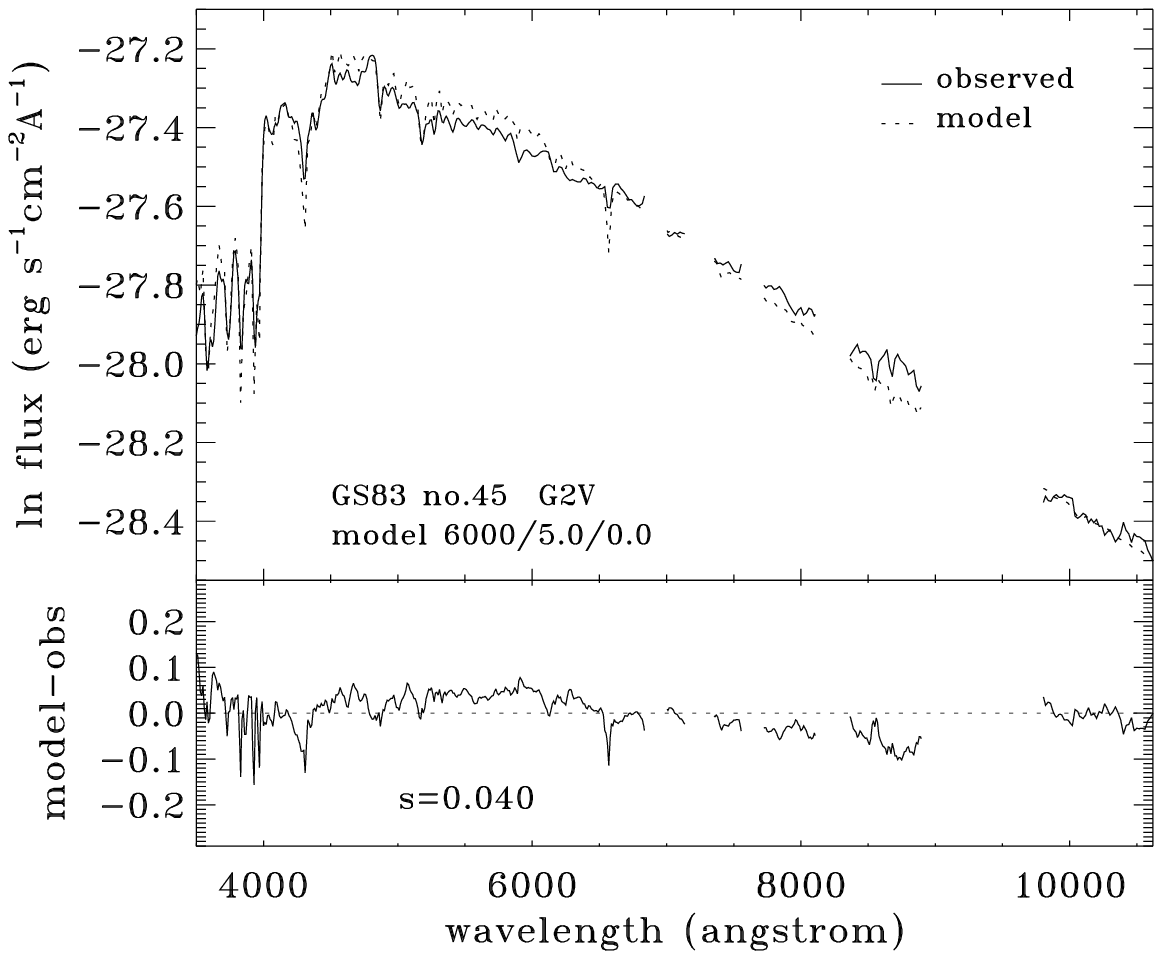} \\
\end{tabular}
\end{center}
\caption{The observed SED of star no.\ 45 from GS83 (namely, HD 154760, of
spectral type G2\,V) is compared with several ATLAS models. Left plots are for
fixed gravity and $\Delta \mteff = \pm 250$~K around the reference value of
the central panel. Right plots are instead for fixed  \teff\ and $\Delta
\mlogg = \pm 0.5$~dex. At the bottom of each panel we display the residual
function $X_{(i,j)}(\lambda)$ according to eq.~(\ref{eq:difflnfluxes}) with
its standard deviation, $s$, from eq.~(\ref{eq:stdev}) as labelled. The central
panel is the ATLAS best-fit solution for this star.}
\label{fig:comparison}
\end{figure*}

\begin{figure*}[!t]
\centerline{
\psfig{file=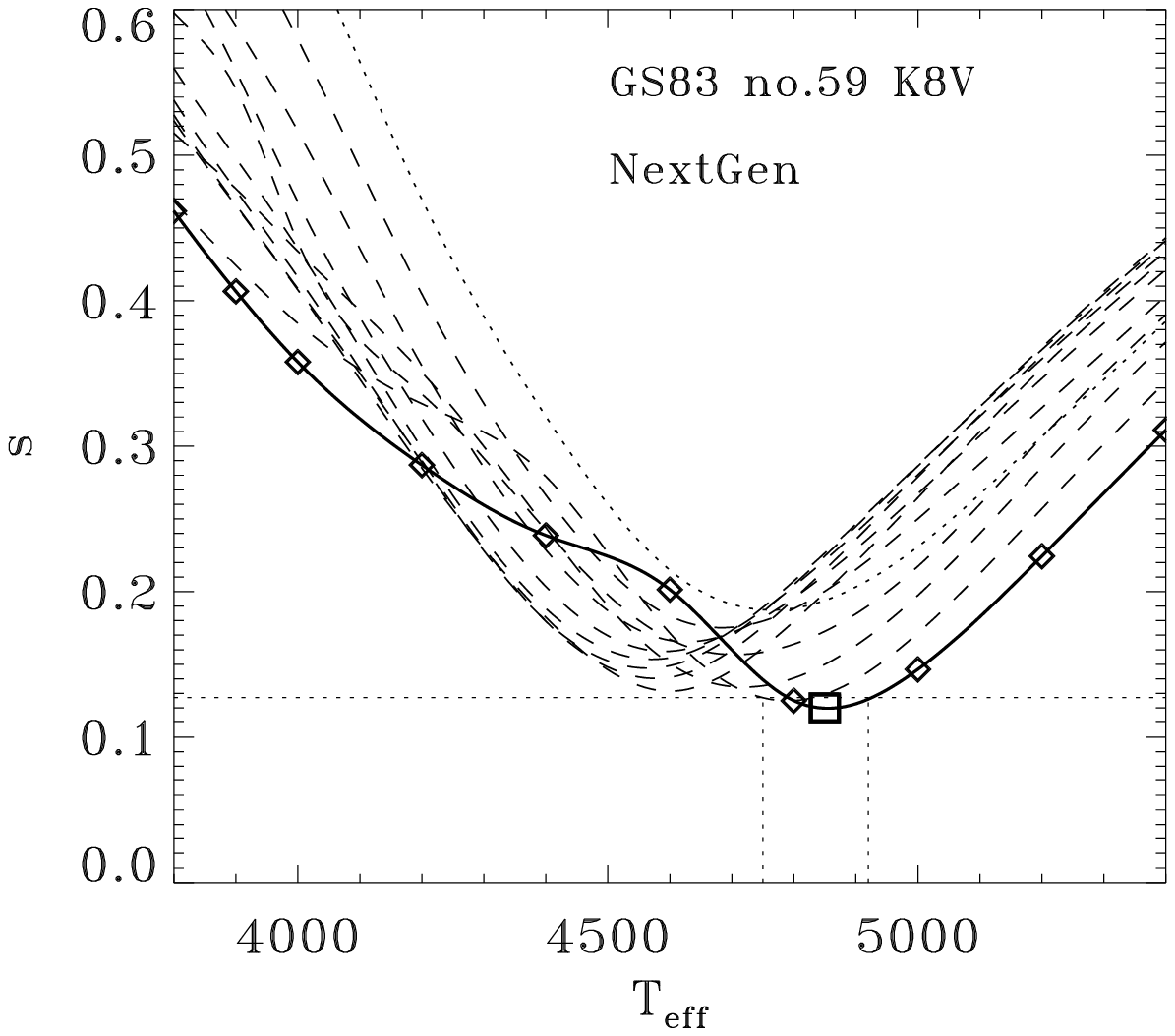,width=0.35\hsize} 
\psfig{file=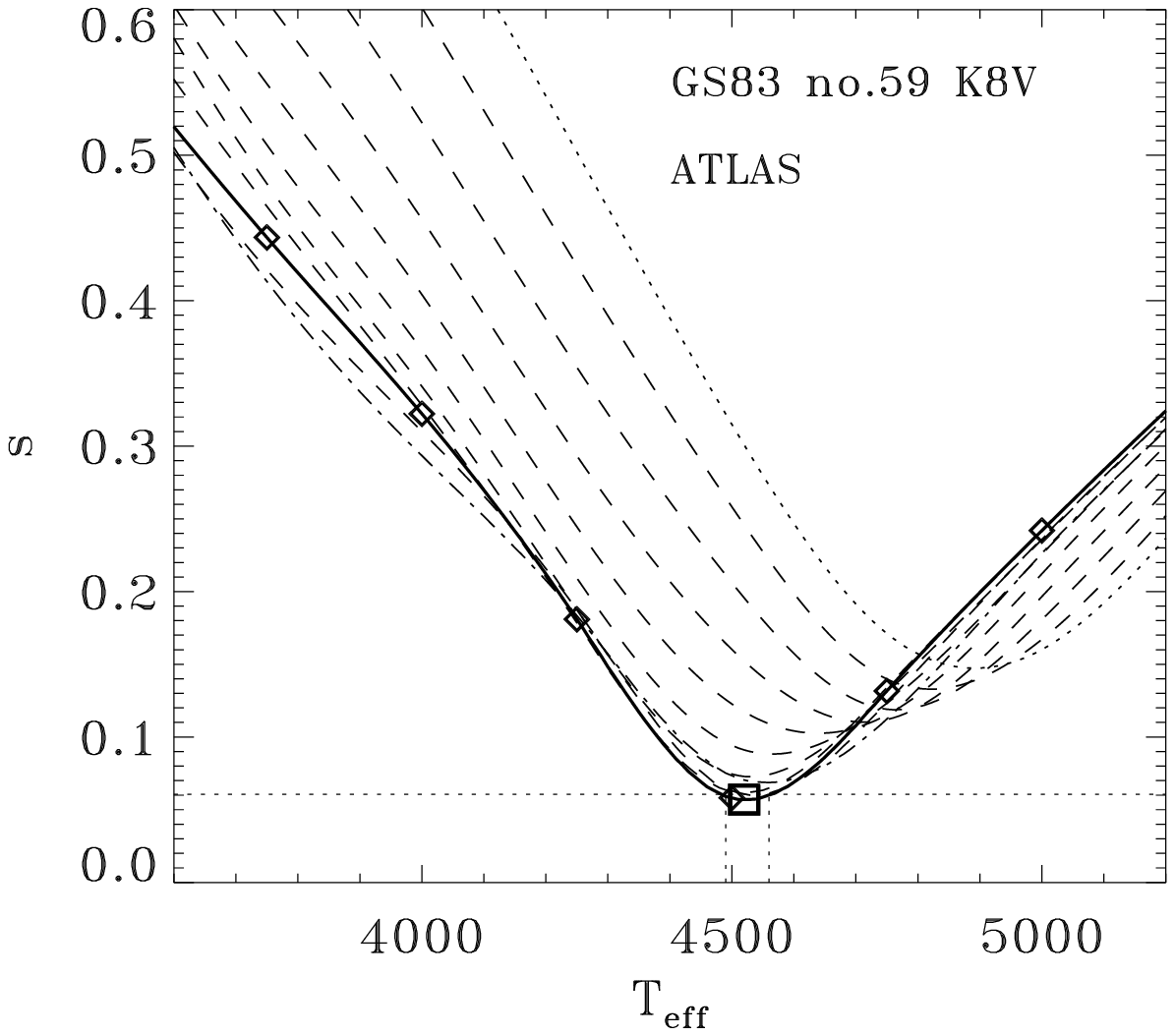,width=0.35\hsize}
}
\caption{Standard deviation of flux residuals for NextGen (left panel) and
ATLAS (right panel) fitting models for star BD\,+38\,2457. Each curve connects
equi-gravity points in the grid vs. \teff\ (dotted line for \logg=0,
dash-dotted for $\mlogg_\mathrm{\scriptscriptstyle NG}=5.5$ or
$\mlogg_\mathrm{\scriptscriptstyle ATLAS}=5.0$, dashed for intermediate
values). Solid line is the best-gravity solution with its minimum marked by
the big open square. The horizontal dotted line shows the upper limit for
$s_\mathrm{min}$ at a 2-$\sigma$ confidence level, as resulting from an $F$
statistical test. The fiducial fundamental parameters for this case are
\teff$=4850^{+70}_{-100}$~K, \logg=$5.5\pm 0.5$~dex with the NextGen models,
and \teff$=4520^{+40}_{-30}$~K and \logg=4.0$\pm 0.5$~dex  with the ATLAS
models. Note the better accuracy of the ATLAS fit ($s_\mathrm{min} = 0.06$)
compared to NextGen ($s_\mathrm{min} = 0.12$).}
\label{fig:spline.gs59}
\end{figure*}

Operationally, the spectrum of the {\it i-th} target star is compared with the
{\it j-th} synthetic SED along the common wavelength range, deriving a
residual function
\begin{equation}
X_{(i,j)}(\lambda) = \ln f_i(\lambda) - \ln f_j(\lambda) +{\rm k}
\label{eq:difflnfluxes}
\end{equation}
in the flux logarithm domain, as shown in Fig.~\ref{fig:comparison}.
The offset constant, k, in eq.~(\ref{eq:difflnfluxes}) is such as
\begin{equation}
\sum_\lambda X_{(i,j)}(\lambda) = 0,
\label{eq:offset}
\end{equation}
so that 
\begin{equation}
{\rm k} = < [\ln f_j(\lambda) - \ln f_i(\lambda)]>,
\label{eq:koffset}
\end{equation}
while the standard deviation 
\begin{equation}
s(X)_{(i,j)} = \sqrt{{\it Var[X(\lambda)]}}
\label{eq:stdev}
\end{equation}
provides a measure of the spectral likelihood between observations and
theoretical models.\footnote{The freedom degrees for $s(X)$ are settled by the
number $N$ of wavelength points after spectrum broadening, as described in
previous section.} The underlying hypothesis of our approach is of course that
a non-degenerate trend exists for the $s(X)$ function in the (\teff, \logg)
phase space, so that a univocal best solution can be found for a given
input SED. For each target star we then mapped the $s(X)$ distribution by
matching the whole grid of synthetic model atmospheres, and searched for an
absolute minimum, $s_{min}$, after performing a cubic spline
interpolation of the $s(X)$ points at each gravity. This allowed
us to locate the best-fitting values of \teff\ and \logg\  with a nominal
resolution of $\Delta \mteff \sim 10~{\rm K}$ and $\Delta \mlogg \sim 0.5~{\rm
dex}$, respectively.

Statistical uncertainty of fiducial distinctive parameters was estimated by
means of a one-tail $F$ test on the value of $s_{min}$, at a 95\% confidence
level.\footnote{The freedom degrees of the $F$ distribution, in this case are
simply $N-1$, cf.\ previous footnote.} The $s_{min}$ confidence interval was
translated into an equivalent ($\Delta \mteff$, $\Delta \mlogg$) error box
relying on a first-order estimate of $\partial s(X) / \partial \mteff$ and
$\partial s(X) / \partial \mlogg$ evaluated around the $s_{min}$ region in the
phase space. 
An example of the fitting procedure for a star in the GS83 sample is
displayed in Fig.~\ref{fig:spline.gs59}.

The robustness of our minimization procedure was probed by means of
a bootstrap test. We added a 10\% noise to the full set of Kurucz synthetic 
SEDs and tried our best fit to recover the original
(i.e.\ unperturbed) reference parameters. In all cases, the correct \teff\ was
identified, typically with a 1--2\% uncertainty, a value that raised to 
5--8\% just for the few poorest cases.
As far as surface gravity is concerned, the nominal values of the reference
models were picked up within a $\pm 0.5$~dex in 98\% of the cases
(i.e. with only 8 outliers out of 409 fitted SEDs).

\section {Temperature scale calibration}
\label{sec:teffbc}

Out of the total of 334 stars in the GS83 and JHC84 libraries, a consistent
fitting solution for the ($\mteff, \mlogg$) fundamental parameters was found
for 272 and 230 stars, respectively, using ATLAS and NextGen reference
grids. Most of the remaining unfitted objects are M and O-B stars, that is at
the two extreme edges of the temperature scale, where a fair value for
$s_{min}$ cannot be confidently located within the theoretical model grid. The
NextGen code, however, has proven to be marginally more efficient in the fit
of M stars (of a total of 38 stars in this class, 10 were successfully matched
by ATLAS and 24 by NextGen).

\begin{figure}[!t]
\psfig{file=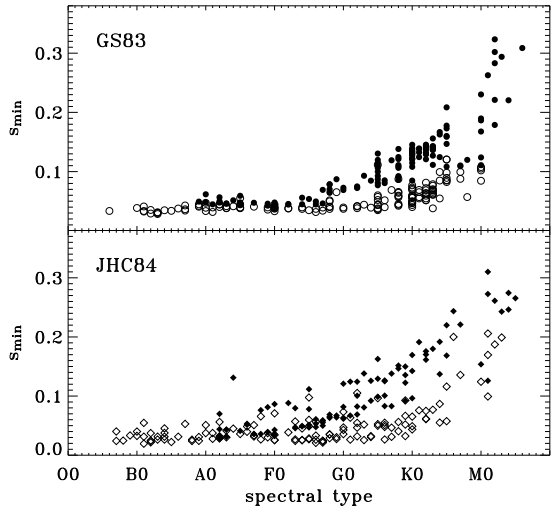,width=\hsize}
\caption{The $s_\mathrm{min}$ distribution vs.\ spectral type for the stars of
the GS83 (upper panel) and JHC84 atlas (lower panel). Open markers
indicate the results for ATLAS models, filled symbols those from the NextGen
grid. Note, for the latter, that a limit at  \teff $\le 10\,000$~K does not
allow any fit to O--B stars. The value of $s_\mathrm{min}$ is a measure of the
mean percent accuracy of the best fit to the observed spectrum.}
\label{fig:sdevmin_teff}
\end{figure}

The accuracy of the ATLAS and NextGen model libraries in the fit of GS83 and
JHC84 stars can be analysed by means of Fig.~\ref{fig:sdevmin_teff}. In the
two panels of the figure we report the distribution of the residual standard
deviation of the best fits for stars in the two observed samples. As a common
feature in the two plots, note that NextGen provides in general a poorer fit
compared to the ATLAS code. This is particularly evident for G and K stars
($\mteff \sim 5500 \to 4000$~K), where ATLAS is a factor of two better than
NextGen in terms of best-fitting variance.
The figure also shows that the accuracy in the definition of the temperature
scale directly depends on the wavelength baseline of the spectra. Compared
with the GS83 stars, in fact the JHC84 fits are slightly poorer, given a
narrower spectral range for the JHC84 library (i.e.\ $\Delta \lambda \sim
4000$~\AA\ vs.\ 7500~\AA\ for GS83).

The incomplete treatment of molecular opacity in the Kurucz code is well
evidenced in both plots of Fig.~\ref{fig:sdevmin_teff}, facing the sharp
increase of standard deviation in the fit of late-K and M stars. The same
problem seems to affect, at a similar level,
also the NextGen fits confirming in any case a still unsolved and pervasive
discrepancy of the theory to self-consistently reproduce cool stars.

\begin{figure*}[!ht]
\centerline{
\psfig{file=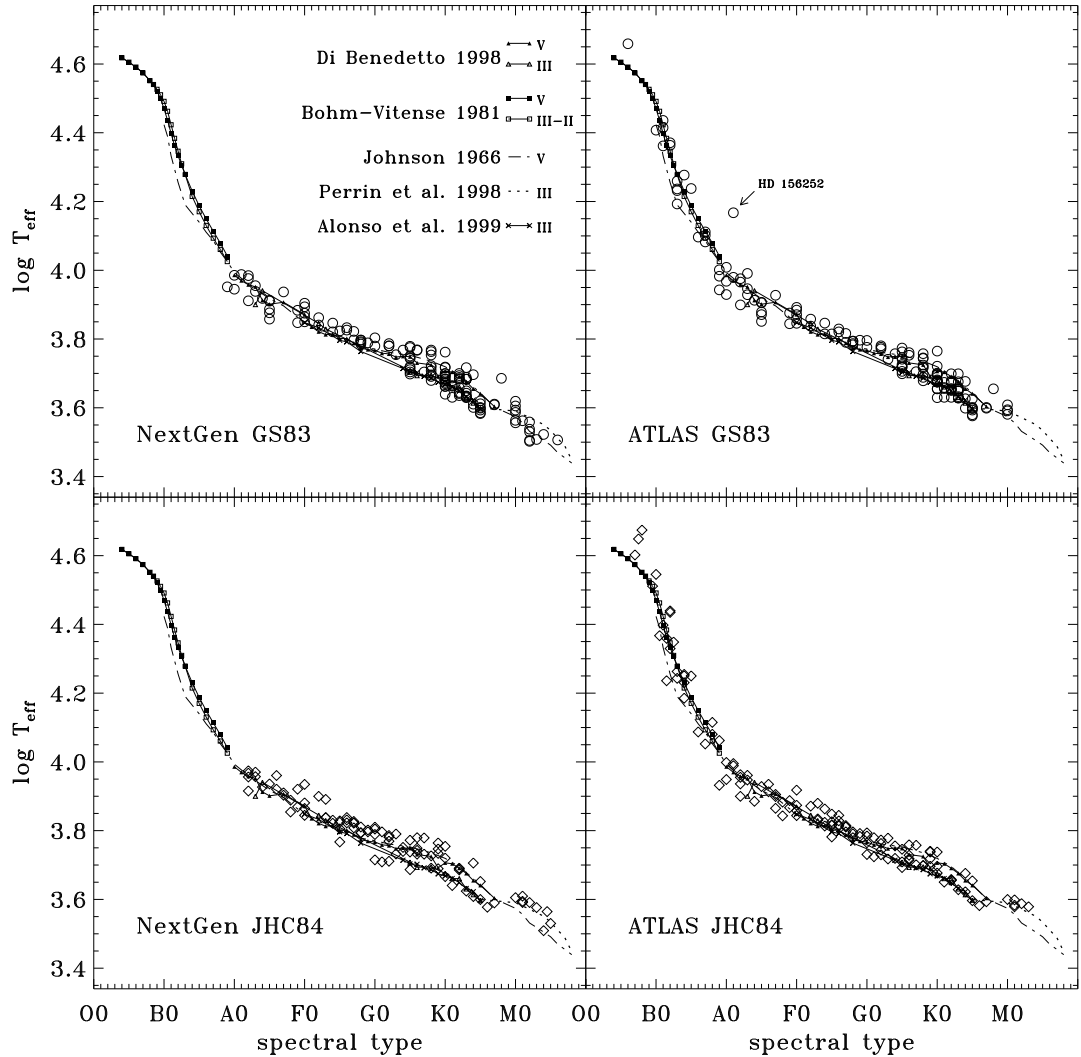,width=0.9\hsize}
}
\caption{The effective temperature scale as derived from the GS83 (upper
panels; dot markers) and JHC84 stars (lower panels; diamond markers), after
NextGen (left panels) and ATLAS (right panels) fits. Our results are compared
with several empirical calibrations from the literature, as indicated in the
top left panel. The outlier star HD 156252 is indicated in the top right panel
(note that this object is not comprised in the NextGen plots due to an upper
limit to effective temperature of the model grid at \teff~$\le
10\,000$~K). See text for discussion.}
\label{fig:teffscale}
\end{figure*}

The temperature scale resulting from the ATLAS and NextGen fits is displayed
in Fig.~\ref{fig:teffscale}. We compared with a number of empirical mean loci
for dwarfs and giants including the \citet{johnson66} classical compilation
and the \citet{bohm81} scales for hot stars. We also considered the recent
calibration of F0--K5 giant stars from \citet{alonso99} and the extension to
late-M giants of \citet{perrin98}, which includes the \citet{ridgway80}
data. Furthermore, \citet{dibenedetto98} provided accurate and systematic
measures of the effective temperature for a wide sample of 537 stars of A to K
spectral type, within an internal accuracy of $\pm 1$\% in the individual
\teff\ estimates, using the surface brightness technique
\citep{wesselink69}, calibrated by means of the angular diameters of 22 stars. 
His mean locus for the dwarf and giant subsamples is
superposed to our data in Fig.~\ref{fig:teffscale}.

In general, we find a consistent trend  between the \teff\ scale from the 
theoretical fits of the GS83 and JHC84 stars and the empirical reference
calibrations. A more scattered distribution of our points derives, of course,
from the fact that these are fits of individual stars, instead of a mean
locus, and because of the luminosity and metallicity spread of the sample.
Only one clear outlier appears among the GS83 stars in the top-right panel of 
Fig.~\ref{fig:teffscale}; this is star HD 156252 (alias 38 Oph), classified by
GS83 as a type A1V star with (dereddened) (B-V) = $-0.14$ mag and color excess
E(B-V) = 0.16 mag. The exceedingly blue color calls for a warmer fitting
temperature ($\mteff \sim 14\,000$~K) with ATLAS, while also the NextGen match
suggests a temperature in excess of 10\,000~K, as no minima of the standard
deviation $s$ [see eq.~(\ref{eq:stdev})] were present below that temperature.

This star is reportedly among the most reddened ones in the GS83 list but if
one accounts, alternatively, for a much lower color excess as reported in the
Hipparcos catalog [namely E(B-V) = 0.015 mag], then the evident discrepancy
between fitting temperature and spectral type could easily be recovered.

\begin{figure}[!t]
\begin{center}
\psfig{file=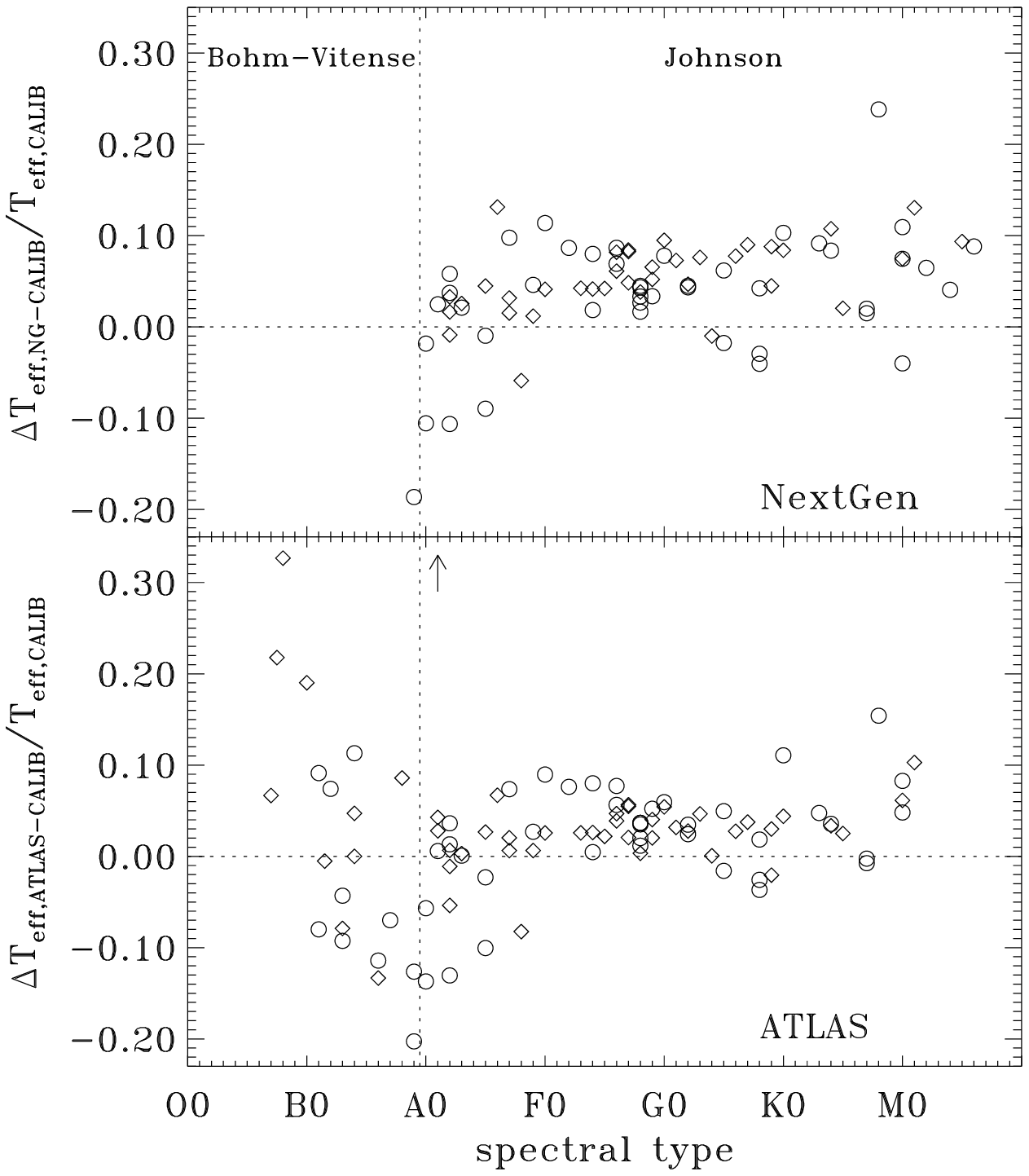,width=\hsize}
\end{center}
\caption{\teff\ residuals vs.\ spectral type of  ATLAS and NextGen best-fits
for the subsample of MK~V stars and the corresponding calibration of
\citet{johnson66}, for A--M stars, and \citet{bohm81} for O--B-types. Open dots
identify GS83 objects, while diamonds mark the JHC84 stars. The outlier
HD\,156252 is off the ATLAS plot, as indicated by the vertical arrow (see text
for discussion).}
\label{fig:diff_johnson}
\end{figure}

\begin{figure}[!ht]
\begin{center}
\psfig{file=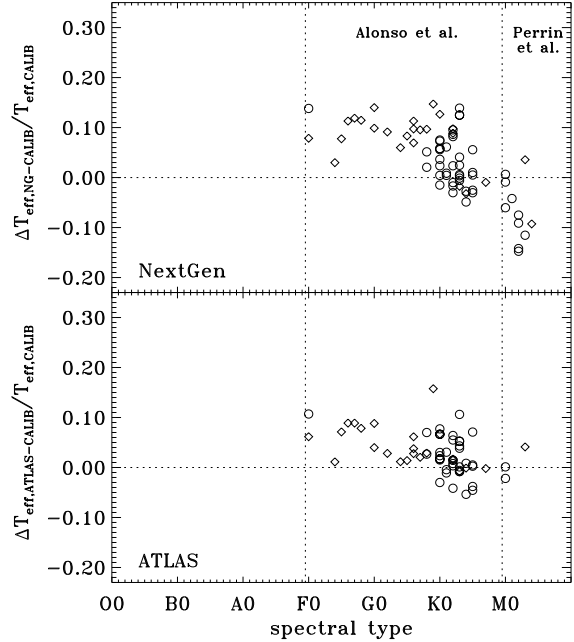,width=\hsize}
\end{center}
\caption{As for Fig.~\ref{fig:diff_johnson}, but for MK~III giant stars vs.\
the corresponding calibrations of Alonso et al.\ (1999) for F--K stars,
and \citet{perrin98} for M-type stars.}
\label{fig:diff_alonso}
\end{figure}

To better single out the differences between the NextGen and ATLAS fits, in
Fig.~\ref{fig:diff_johnson} and \ref{fig:diff_alonso}  we considered
separately the main sequence (MK class V) and giant (MK III) star subsamples
displaying the $\Delta \mteff\ / \mteff$ between our fitting temperature and
the reference calibration of \citet{bohm81} and \citet{johnson66}, for dwarfs,
and Alonso et al.\ (1999) and \citet{perrin98}, for giants.
Fig.~\ref{fig:diff_johnson} shows that both ATLAS and NextGen grids tend to 
fit F to M stars with a 4--8\% warmer effective temperature, the \teff\ excess
being in general higher for the NextGen fits.

The situation is somehow different for giants (see Fig.~\ref{fig:diff_alonso})
with a drift in the point distribution with respect to the Alonso et al.\
(1999) and \citet{perrin98}\ \teff\ calibration (but, again, with the NextGen
output marginally warmer than the ATLAS one).

As for the Kurucz models, a glance to Fig.~\ref{fig:comparison} makes clear
that part of the bias toward higher fitting temperatures might derive from the
blanketing effects in the ultraviolet region of the stellar SED, shortward of
4000~\AA. The residual scatter in this spectral region is in fact a major
source to the global variance when matching observed SED and theoretical
models, thus sensibly constraining the choice of the best-fit solution. More
entangled is the situation for NextGen models, which adopt a 
different chemical mix representative of solar metallicity.
While in fact ATLAS relies on the \citet{andersgrevesse89} solar abundances, 
NextGen assumes the revised values by 
\citet{jaschek95};
for $Z = Z_\odot$, this makes NextGen Fe abundance slightly lower
compared to ATLAS (namely, [Fe$_{\rm NG}$/Fe$_{\rm ATLAS}] \simeq -0.17$~dex). 
However, this feature could hardly explain the observed trend
in the \teff\ calibration as a Fe-poorer model atmosphere should actually 
display a {\it lower} blanketing, thus allowing a {\it cooler} 
temperature to fit the observed SED of stars 
\citep[see, in this sense, the quantitative discussion by][]{buzzonietal01}.\footnote{
As a cross-check in this regard, we tried a fit of the GS83 stars
relying on the Kurucz library with [Fe/H]~$=~-0.1$ instead of solar.
From the operational point of view, this should roughly mimic the NextGen 
solar case. As expected, the GS83 fitting temperatures are in average
50--100~K cooler than the values obtained with the ATLAS library at 
[Fe/H]~=~0.}

\subsection{Bolometric corrections}

Our best-fitting procedure with model atmospheres allows in principle a
straightforward estimate of bolometric luminosity for stars in the GS83 and
JHC84 samples. When coupled with the individual $V$ magnitudes, this could
eventually supply a measure of the bolometric correction ($BC$).

\begin{figure}[!t]
\centerline{
\psfig{file=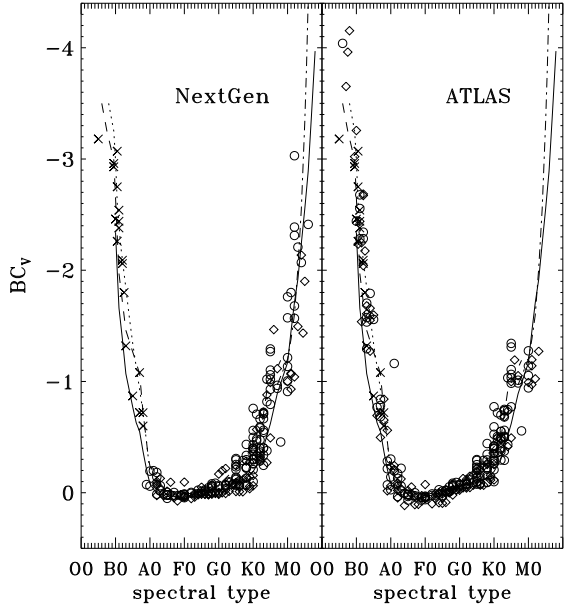,width=\hsize}
}
\caption{Derived bolometric correction for GS83 (open dots) and JHC84
(diamonds) stars according to ATLAS and NextGen model atmospheres. Our results
are compared with the empirical calibrations of Flower (1977, dotted line for
MS stars, dashed for giants), Johnson (1966, solid), Bessell (1991,
dashed-dotted), and Code et al.\ (1976, cross symbols).}
\label{fig:bc}
\end{figure}

In order to set the $BC$ scale, for our calculations we identified the
theoretical template for the Sun as the ATLAS model SUNK94, with
(\teff, \logg, [M/H]) = (5777~K, 4.44, 0.0) according to \citet{castellietal97}. 
Its theoretical SED has been convolved with the $V$ filter
profile of \citet{bessell90} and photometric zero points were tuned up such as
to  have $BC_\odot = Bol_\odot - V_\odot = -0.07$ \citep{bcp98}.

The $BC$ for each star in the GS83 and JHC84 samples is then computed as:
\begin{equation}
BC = -2.5 \log{(\sigma \mteff^4)} - V  -{\rm k} + 2.22.
\label{eq:bc}
\end{equation}
Note that in eq.~(\ref{eq:bc}), the $V$ magnitude derives from the convolution
of the {\it observed} spectrum while the offset ``k'' (that properly scales the
bolometric magnitude of the theoretical fitting SED) is from
eq.~(\ref{eq:koffset}).

Our results are compared, in Fig.~\ref{fig:bc}, with other standard
calibrations for dwarf and giant stars vs.\ spectral type. We considered in
particular the work of \citet{flower77} and \citet{johnson66} and its later
revisions of \citet{bessell91} (for late-K and M dwarf stars), and
\citet{code76} (for hot O--B stars). When necessary, bolometric scales were
shifted consistently such as to assure $BC_\odot = -0.07$.

The final reference calibration for \teff\ and $BC$ vs. spectral type for
dwarf and giant stars according to ATLAS and NextGen is summarized in
Table~\ref{tab:bc}. Given the limited spectral coverage of the observed
spectra, our bolometric extrapolation suffers, of course, from intrinsic
uncertainties at the two extremes edges of the temperature scale, where a
substantial fraction of stellar energy is emitted outside the optical range.
The $BC$ calibration of Table~\ref{tab:bc} for B and M stars should
therefore be taken with some caution since it critically relies on the
theoretical input physics.

\begin{figure*}[!t]
\centerline{
\psfig{file=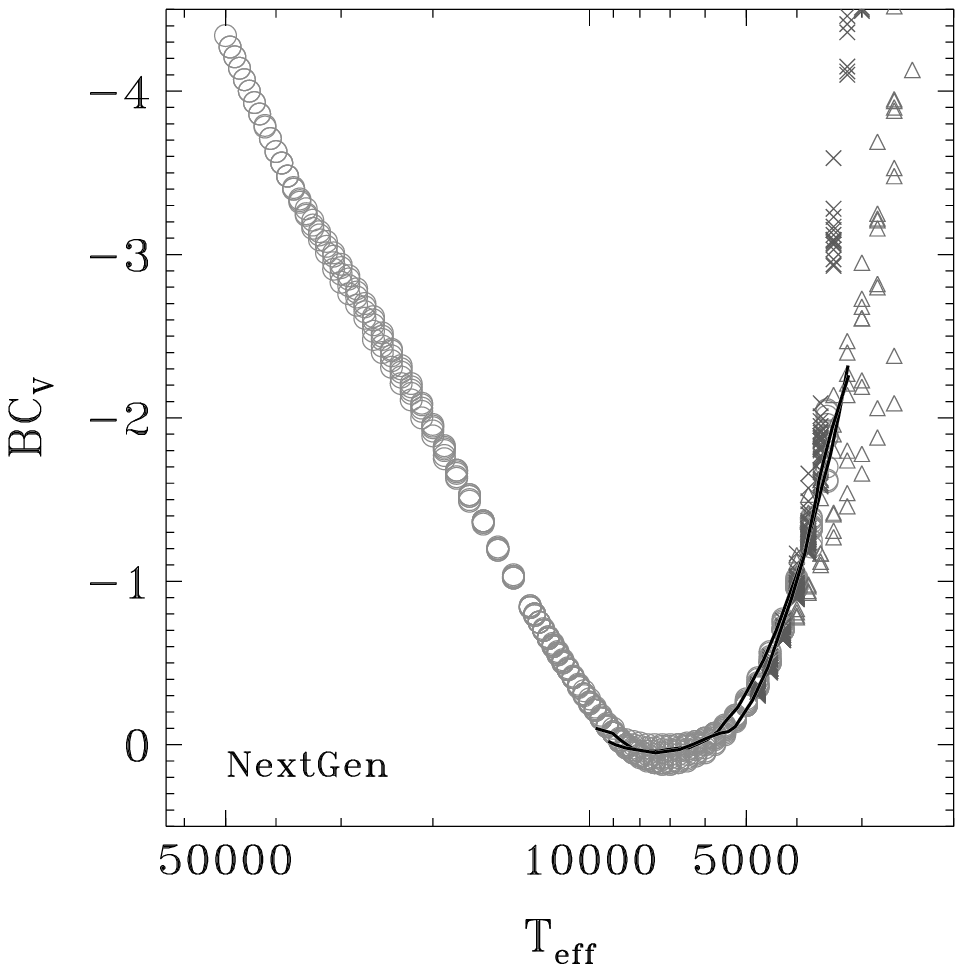,width=0.45\hsize}
\psfig{file=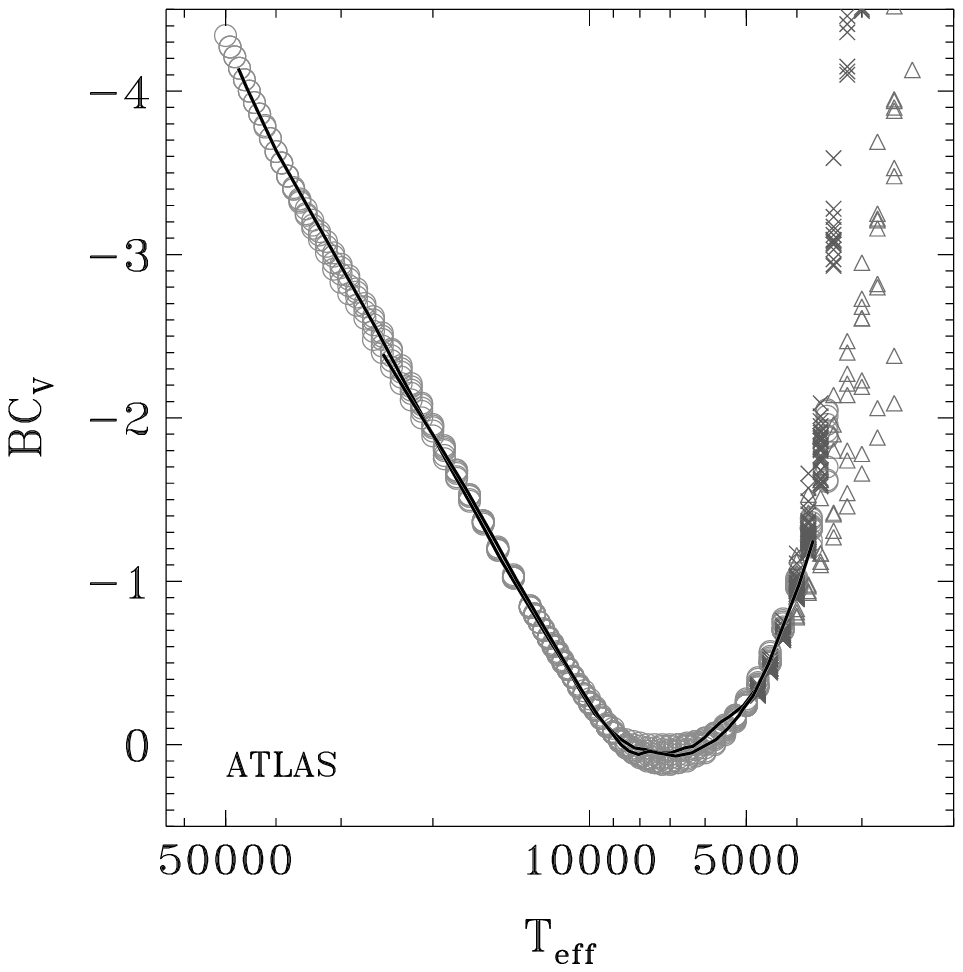,width=0.45\hsize}
}
\caption{As for Fig.~\ref{fig:bc}, but vs.\ temperature scale for
several theoretical calibrations. Our results, from Table 1 (solid
lines), are compared with those of \citet{bcp98}, also based on the
ATLAS models (open dots), \citet{plez92} for cool giant models
(crosses), and \citet{edv93} for M dwarfs (triangles), both relying on
the NMARCS atmosphere code of \citet{gus03}. Left panel reports our
NextGen calibration, while right panel is for the ATLAS theoretical
locus.}
\label{fig:bcp}
\end{figure*}

A more extensive discussion of this issue, from a fully theoretical point of
view, has been carried out by \citet{bcp98} based on the
ATLAS and NMARCS model predictions, the latter ones as from the calibrations
of \citet{plez92} for giants and \citet{edv93} for dwarfs. A comparison of our
results with those of \citet{bcp98} is shown in Fig.~\ref{fig:bcp} comforting,
however, on the general agreement of our output.

\begin{deluxetable}{crrrrrrrrr}

\tablecolumns{10} 

\tablewidth{0pc} 

\tablecaption{Fiducial ATLAS and NextGen calibration for temperature scale and
bolometric correction}

\tablehead{\colhead{~} & \multicolumn{4}{c}{ATLAS} & \colhead{~} & \multicolumn{4}{c}{NextGen} \\

\colhead{~} & \multicolumn{2}{c}{Dwarfs} & \multicolumn{2}{c}{Giants} & \colhead{~} & \multicolumn{2}{c}{Dwarfs} & \multicolumn{2}{c}{Giants} \\

\cline{2-5} \cline{7-10}\\

\colhead{Sp. Type} & \colhead{\teff} & \colhead{BC} & \colhead{\teff} & \colhead{BC}  & \colhead{~} & \colhead{\teff} & \colhead{BC} & \colhead{\teff} & \colhead{BC} }

\startdata 
B0   &  30430 & -2.99 &  28640 &       & &        &       &        &       \\
B1   &  25830 & -2.56 &  24910 & -2.39 & &        &       &        &       \\
B2   &  22000 & -2.14 &  22020 & -2.11 & &        &       &        &       \\
B3   &  18920 & -1.75 &  19470 & -1.84 & &        &       &        &       \\
B4   &  16490 & -1.41 &  17230 & -1.56 & &        &       &        &       \\
B5   &        &       &  15310 & -1.27 & &        &       &        &       \\
B6   &  13090 & -0.85 &        &       & &        &       &        &       \\
B7   &  11940 & -0.64 &  12330 & -0.74 & &        &       &        &       \\
B8   &  11040 & -0.47 &  11230 & -0.52 & &        &       &        &       \\
B9   &  10330 & -0.31 &  10340 & -0.33 & &        &       &        &       \\
A0   &   9780 & -0.19 &   9640 & -0.18 & &   9730 & -0.10 &        &       \\
A2   &   8980 & -0.07 &   8700 &  0.00 & &   9030 & -0.07 &   9200 & -0.02 \\
A3   &   8680 & -0.03 &   8400 &  0.04 & &   8740 & -0.03 &   8930 &  0.00 \\
A5   &   8200 &  0.02 &   8050 &  0.06 & &   8300 &  0.02 &   8530 &  0.02 \\
A7   &   7810 &  0.03 &   7880 &  0.05 & &   7910 &  0.04 &        &       \\
F0   &   7290 &  0.05 &   7660 &  0.04 & &   7380 &  0.04 &   7810 &  0.04 \\
F2   &   6980 &  0.05 &   7410 &  0.05 & &   7080 &  0.03 &   7470 &  0.05 \\
F5   &   6570 &  0.02 &   6830 &  0.07 & &   6680 &  0.02 &   6730 &  0.03 \\
F7   &   6330 &  0.01 &   6360 &  0.05 & &   6450 &  0.02 &   6350 & -0.00 \\
G0   &   6030 & -0.04 &   5720 & -0.03 & &   6140 & -0.01 &   5850 & -0.05 \\
G2   &   5860 & -0.08 &   5420 & -0.10 & &   5950 & -0.03 &   5620 & -0.07 \\
G5   &   5590 & -0.14 &   5160 & -0.18 & &   5670 & -0.08 &   5410 & -0.08 \\
G7   &   5400 & -0.17 &   5070 & -0.22 & &   5490 & -0.14 &   5250 & -0.11 \\
K0   &   5060 & -0.24 &   4850 & -0.30 & &   5190 & -0.23 &   4870 & -0.27 \\
K2   &   4820 & -0.33 &   4540 & -0.49 & &   4980 & -0.33 &   4550 & -0.47 \\
K5   &   4480 & -0.52 &   3960 & -0.98 & &   4630 & -0.52 &   4090 & -0.90 \\
K7   &   4290 & -0.71 &   3720 & -1.25 & &   4380 & -0.70 &   3860 & -1.17 \\
M0   &   4010 & -0.97 &        &       & &   3990 & -1.05 &   3670 & -1.51 \\
M1   &   3890 & -1.04 &        &       & &   3860 & -1.19 &   3630 & -1.61 \\
M2   &        &       &        &       & &   3720 & -1.37 &   3550 & -1.74 \\
M3   &        &       &        &       & &        &       &   3420 & -1.94 \\
M4   &        &       &        &       & &   3450 & -1.77 &   3180 & -2.26 \\
M5   &        &       &        &       & &   3320 & -2.02 &        &       \\
M6   &        &       &        &       & &   3190 & -2.32 &        &       \\
\enddata
\label{tab:bc}
\end{deluxetable} 

\section{Comparing template SED along the spectral-type sequence}
\label{sec:results}

The whole set of synthetic templates for the GS83 and JHC84 stars prompts a 
straightforward comparison of the ATLAS vs.\ NextGen code performances taking
into account in a self-consistent way the effect of the different input
physics on the match of SED for real stars along the O~$\to$~M spectral-type
sequence. For our test we especially relied on the subset of 216 objects 
from the GS83 and JHC84 catalogs for which a nominally best fit exists both
for the ATLAS and NextGen grids over the temperature range $3500 \leq \mteff
\leq 10\,000$~K.

\begin{figure}[!t]
\psfig{file=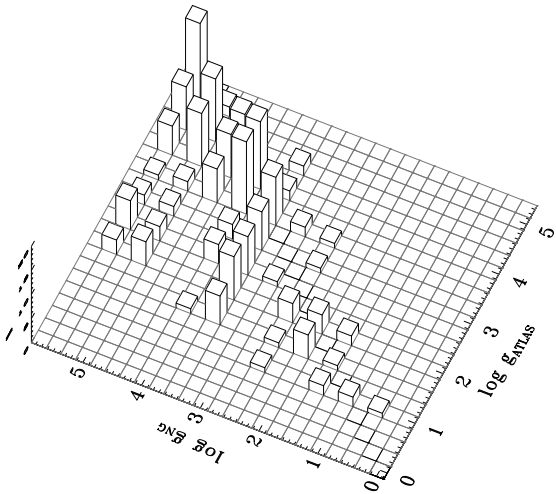,width=\hsize}
\caption{The $\log{g}_{\rm \scriptscriptstyle NG}$ vs. $\log{g}_{\rm
\scriptscriptstyle ATLAS}$ distribution for the sample of 216 stars with
common fitting solutions. The vertical axis shows the frequency number. Note
the excess of high-gravity best-fit solutions for NextGen, compared to the
corresponding ATLAS distribution (see text for full discussion).}
\label{fig:grav_ng_atlas}
\end{figure}

\begin{figure}[!t]
\begin{center}
\psfig{file=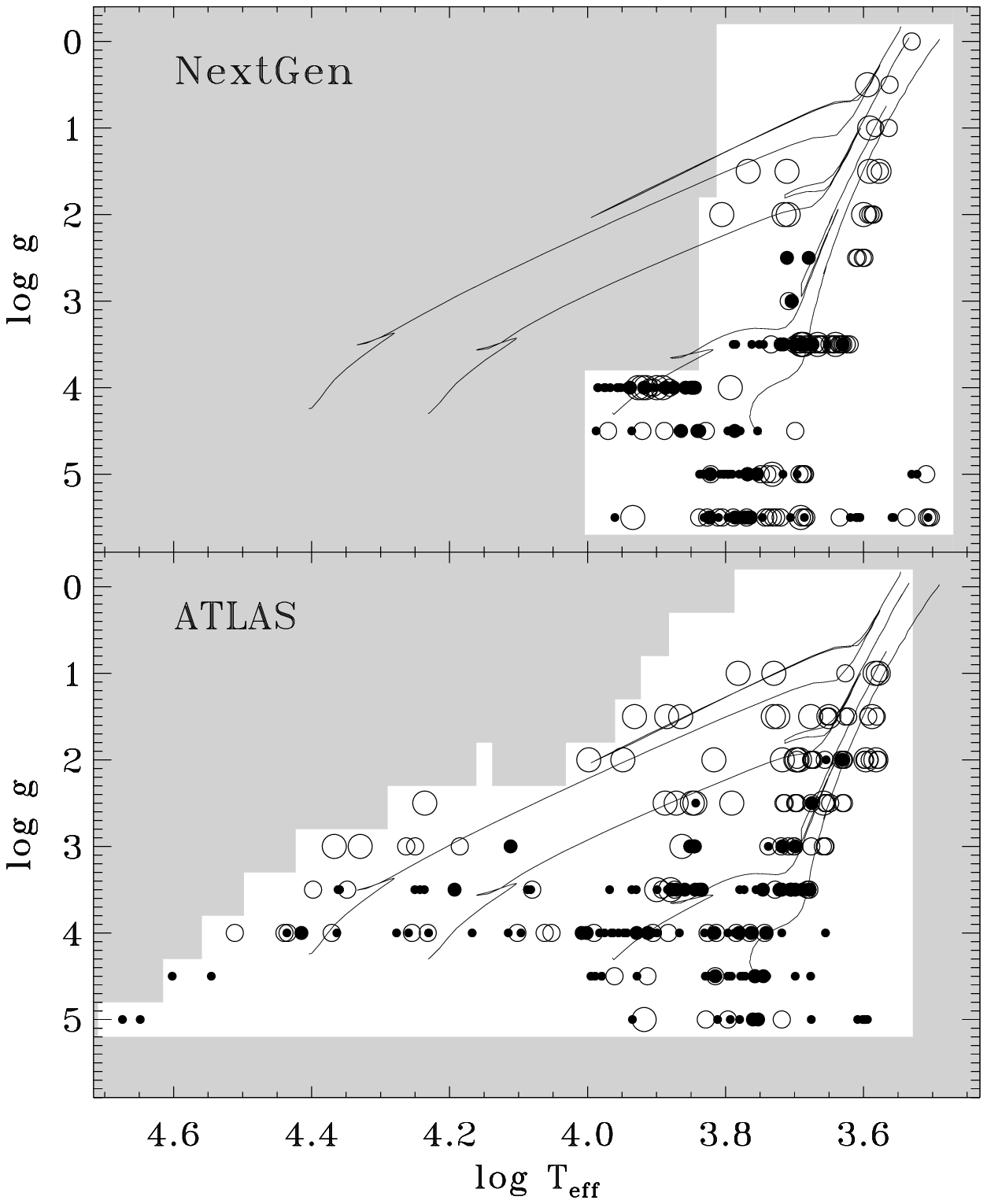,width=\hsize}
\end{center}
\caption{The log \teff\ vs.\ \logg\ distribution of ATLAS and NextGen fitting
output for GS83 and JHC84 stars. The different symbols mark the luminosity
class: MK V~=~$\bullet$, IV~=~big~$\bullet$, III~=~$\circ$,
II--I~=~$\bigcirc$. Overplotted are also the evolutionary tracks for stars of
solar metallicity and $M = 1, 2$, and $5~M_\odot$ from  \citet{girardi00}, and
for $10~M_{\odot}$ from \citet{salasnich00}. The white region show the
parameter space covered by the theoretical libraries (see also
Fig.~\ref{fig:grids}).}
\label{fig:logg_logt}
\end{figure}

A first interesting feature, when comparing the two sets of theoretical
templates, like in Fig.~\ref{fig:grav_ng_atlas}, concerns the distribution of
the fitting gravity. It is evident from the histogram that for an important
fraction of target stars NextGen tends to fit with a higher gravity with
respect to ATLAS. This actually led to a number of ``catastrophic outliers''
among the giant and supergiant stars in the GS83 and JHC84 samples, as shown
in Fig.~\ref{fig:logg_logt}. In the NextGen plot, in fact, (upper panel of the
figure) 23 out of the 100 MK I--III stars are unexpectedly located in the
high-gravity region of the diagram, pertinent to class V dwarfs, with a
nominal ``best-fit'' gravity of $\mlogg =5.5$~dex. Conversely, only four such
gravity outliers are present in the ATLAS diagram (lower panel) with a gravity
of $\mlogg =5$~dex.

\begin{figure}[!t]
\psfig{file=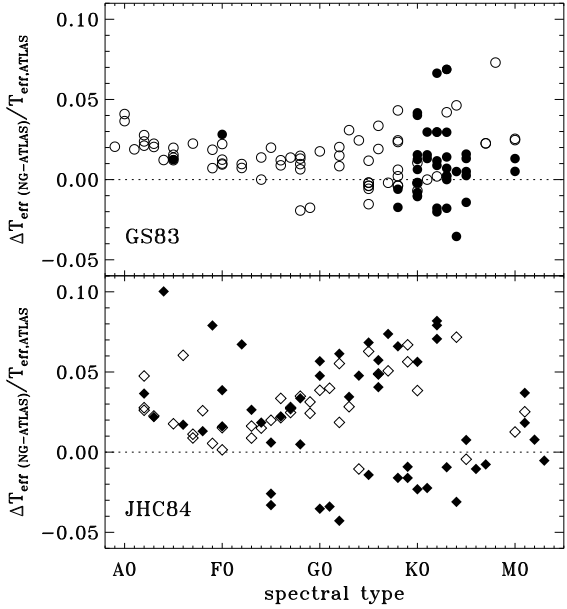,width=\hsize}
\caption{Temperature residuals for the 216 GS83 (upper panel) and JHC84 stars
(lower panel) with both ATLAS and NextGen best-fit solution. Stars are
labelled according to their MK luminosity class (MK IV--V: open markers; MK
I--III: solid markers). Note, in the JHC84 plot, the peculiar distribution of
giant stars, along two distinct point sequences. Over the whole sample,
NextGen tends to predict, in average, an effective temperature about 2\%
warmer than the ATLAS value (see text for a full discussion of these two
important features).}
\label{fig:dteff2}
\end{figure}

A similar trend can also be recognized for the fitting stellar temperature
\teff, as we have been discussing in Sec.~\ref{sec:teffbc}. Again,
Fig.~\ref{fig:dteff2} shows that NextGen \teff\ estimates are in average 2\%
in excess with respect to the ATLAS best-fit values, with a sensibly higher
scatter for the JHC84 stars (see lower panel in the figure), partly depending
on the shorter wavelength baseline compared to the GS83 set of spectra
[$\sigma(\Delta \mteff / \mteff) = 0.031$~dex for the JHC84 sample vs.\ a
value of 0.018~dex for the GS83 stars].

\begin{figure}[!t]
\centerline{
\begin{tabular}{c}
\psfig{file=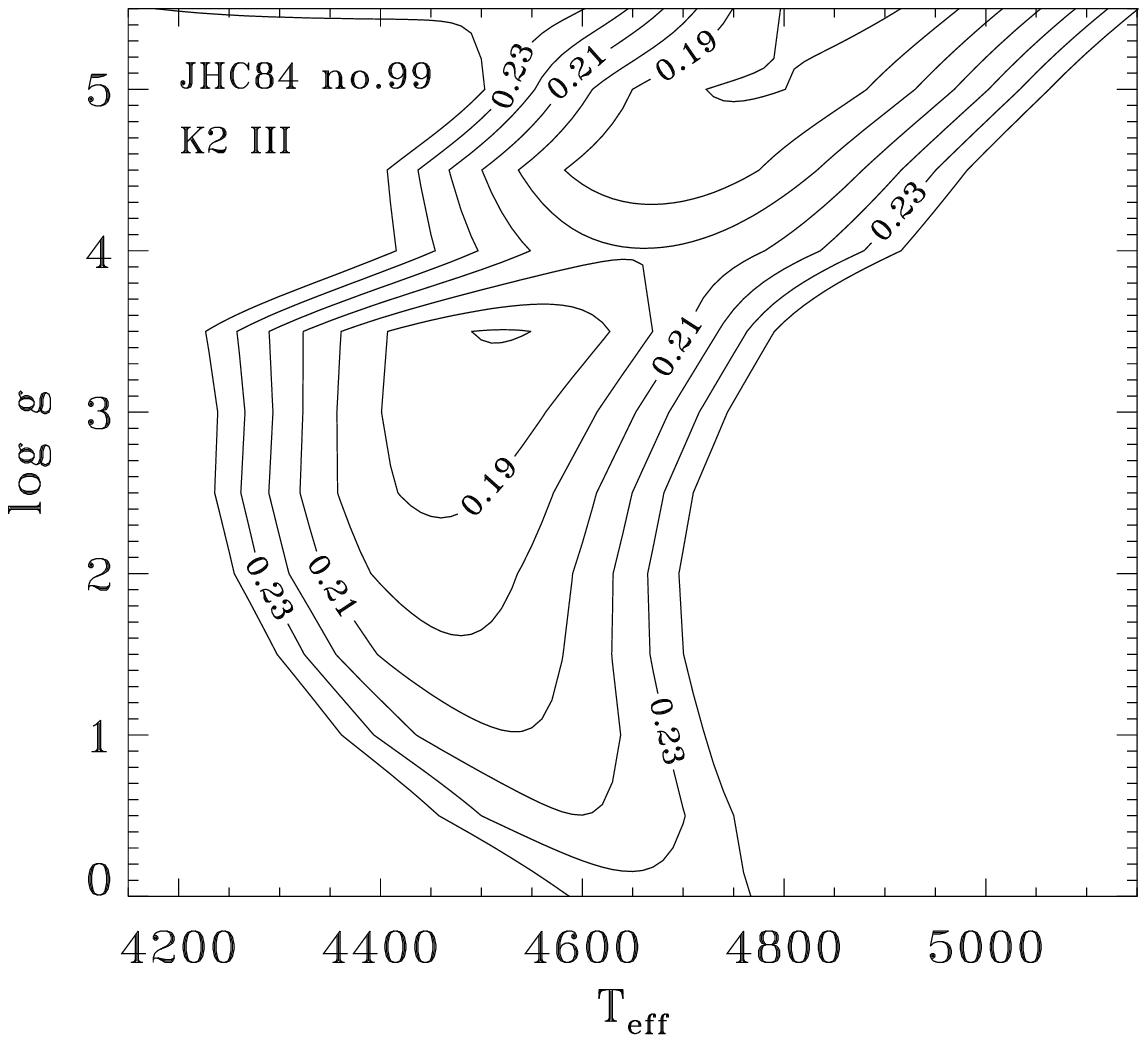,width=0.85\hsize} \\
\psfig{file=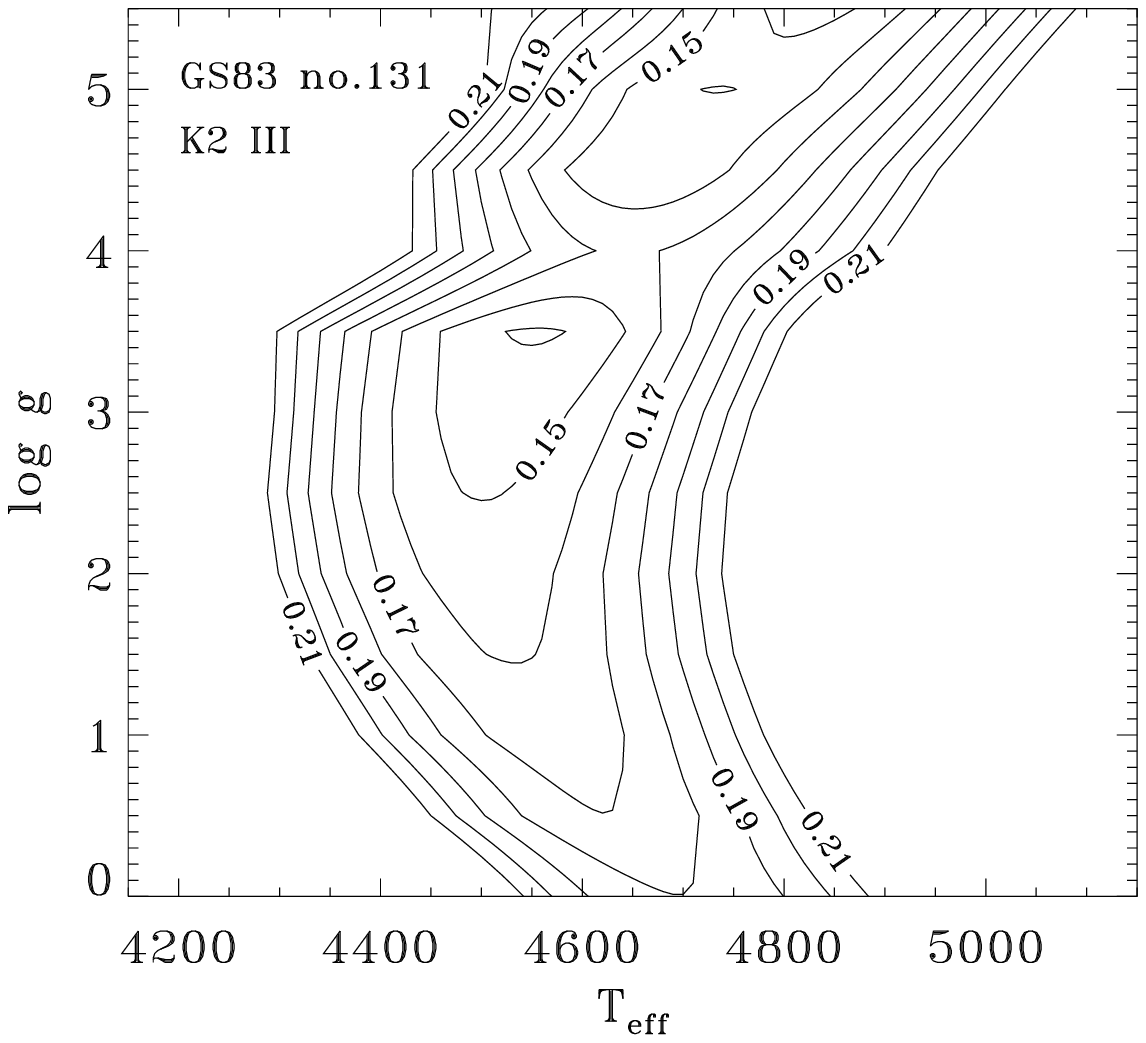,width=0.85\hsize}
\end{tabular}
}
\caption{An illustrative example of the NextGen fitting procedure for two K2
giant stars from the GS83 and JHC84 samples. Plotted is the map of the
standard deviation, $s$, of residual flux between observed and theoretical SED
across the model grid [according to eq.~(\ref{eq:stdev})] in the \teff\ vs.\
\logg\ phase space. One sees that two best-fit solutions can be identified in
each plot, one correctly placing the K2 III stars in the low-temperature
low-gravity region (i.e.\ $\mteff \sim 4500$~K, $\mlogg \sim 3.0$~dex) and the
other {\it nominally better} one shifting stars to slightly  warmer
temperature and much higher gravity (\teff~$\sim 4800$~K, \logg~$\sim
5.5$~dex).}
\label{fig:spline.jhc99.gs131}
\end{figure}

The tendency of NextGen to overestimate temperature and gravity can be
illustrated by means of Fig.~\ref{fig:spline.jhc99.gs131}, where we map the
distribution of the fit variance ($s$) across the theoretical grid for two MK
III giants in the GS83 and JHC84 samples.\footnote{This plot is basically
a projected view of the 3-D fitting surface, like that shown in
Fig.~\ref{fig:spline.gs59}} One sees from the plots that actually {\it two}
physically distinct solutions exist for these stars, one that correctly
locates both K2III giants in the low-temperature low-gravity range [namely,
(\teff, \logg)~$\sim$~(4500~K, 3.0~dex) in our example] and adopts a spherical
model (i.e.\ in the $\mlogg \le 3.5$ domain), and the other 
({\it nominally better}) one that assumes a plane-parallel geometry but places 
stars at a much higher $\mlogg \sim 5.5$~dex and $\mteff \sim 4800$~K.
This apparent ``bimodality'' in the solution space is also well evident in the
JHC84 panel of Fig.~\ref{fig:dteff2}, where giant star distribution appears to
split in two distinct sequences depending whether the spherical or the
plane-parallel solution prevails as a best fit.\footnote{The same effect is
not equally evident in the GS83 plot due to the dominant fraction of dwarf
stars in this sample.}

\begin{figure}[!t]
\psfig{file=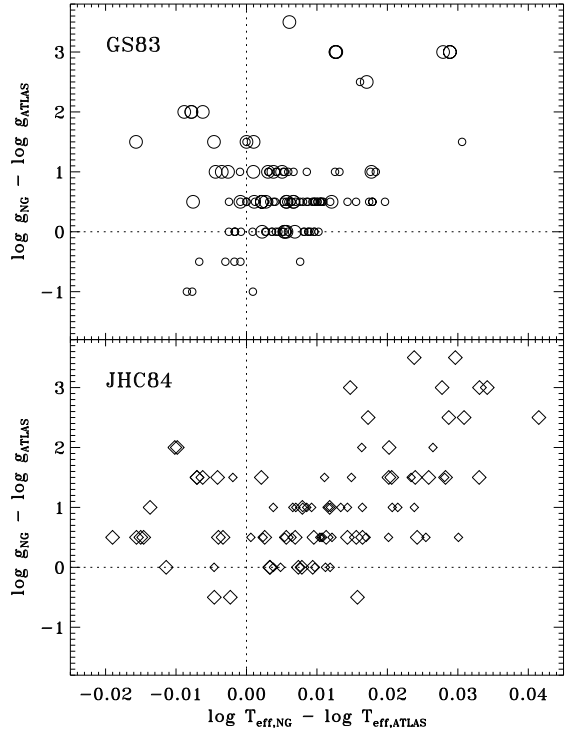,width=\hsize}
\caption{Temperature and gravity difference between ATLAS and NextGen best-fit
solutions for 216 stars in the GS83 and JHC84 samples. Marker size is
proportional to the MK luminosity class (i.e.\ big markers = MK I--III giants,
small markers = MK IV--V dwarfs). It is evident a correlation between $\Delta
\mteff$ and $\Delta \mlogg$, especially for giant stars in the JHC84 plot.}
\label{fig:teff_grav_ng_atlas}
\end{figure}

Note, by the way, that some correlation in the temperature and gravity excess,
when fitting empirical SED with theoretical models, can naturally be expected
on the basis of the arguments pointed out by \citet{buzzonietal01}. Their
experiments showed in fact that a correspondingly higher gravity should likely
be required to recover, at medium-high resolution, ``too shallow'' absorption
features as predicted by a too warmer model forced to match the low-resolution
SED of a given star. According to \citet{buzzonietal01}, such a tight
dependence between $\Delta \log \mteff$ and $\Delta \mlogg$ can be written in
the form
\begin{equation}
{{\Delta\mlogg}\over{\Delta \log \mteff}} = 3000\, \left({1000\over \mteff}\right)^3 \; \mathrm{dex\;K}^{-1}.
\label{eq:2}
\end{equation}
A consistent trend in this sense is confirmed in
Fig.~\ref{fig:teff_grav_ng_atlas}, that reports the temperature and gravity
differences between ATLAS and NextGen fiducial solutions for the 216 target
stars in common.

\subsection{Sphericity effects on theoretical SED}

Such a different behaviour of ATLAS and NextGen model output directly calls,
of course, for a distinct physical approach in the calculation of the inner
structure of the stellar atmosphere. This is especially true for giant stars,
where the plane-parallel model atmospheres of ATLAS are to be compared with
the spherical-shell geometry of NextGen. The impact of geometry on the
emerging flux of the theoretical models has been first assessed in a
pioneering work by \citet{st84} on the atmospheres of M and C stars, and more
extensively explored in the recent years by Plez and collaborators
\citep{plez90,plez92}.

Basically two intervening effects modulate the integrated SED of spherical
models with respect to their corresponding plane-parallel cases. First, as a
general trend for fixed \teff\ and \logg, spherical model atmospheres tend to
display a lower electronic pressure and a cooler temperature profile vs.\
stellar spatial coordinate \citep[i.e.\ radius or optical depth,
cf. e.g.][]{st84}. To some extent, this is the physical consequence of the
decreasing gravity when moving outward of stellar photosphere; with a lower
gravity, in fact, thermodynamical equilibrium in the external layers readjusts
such as to allow a lower  pressure of the electronic plasma (because of an
increased mean distance between atoms and a higher dumping potential for the
bound-bound and bound-free $e^-$ transitions) and a cooler temperature, still
sufficient however to ``sustain'' the atmosphere structure.

As a result, for fixed \teff\ and \logg, the SED of a ``spherical'' star is
therefore expected to display sharper absorption lines and a ``redder''
continuum. Among others, this should also reflect in a less severe blanketing
absorption \citep[see in this sense the experiments of][]{hau99b} as a
consequence of a reduced blend of metal absorption lines at short wavelength.

A second related effect that should be dealt with, when comparing
plane-parallel and spherical model atmospheres, concerns limb darkening. Due
to the geometry, in fact, the integrated flux that emerges from a
``spherical'' star receives a more important contribution from low-gravity
cooler layers, and appears therefore in average ``cooler'' with respect to its
corresponding plane-parallel model \citep{claret03}.

This feature also emerges from the recent results of
\citet{fields03} on the microlensing surface scanning of the K3 giant star
related to the EROS BLG2000-5 event. Surface brightness measurements for this
star are in fact inconsistent with the NextGen best-fit predictions at higher
than 10$\sigma$, and indicate that the derived $T(\tau)$ vertical structure of
the theoretical atmosphere sensibly overestimates limb-darkening effects.


\section{Summary and conclusions.}
\label{sec:conclusions}

In this work we carried out a combined comparison of the two theoretical codes
ATLAS \citep{kurucz92b} and NextGen \citep{hau99a,hau99b} for stellar
atmosphere synthesis. Our tests relied on the fit of a set 334 target stars of
nearly solar metallicity, spanning the whole sequence  of spectral types and
luminosity class, observed in the optical range by \citet{gs83} and
\citet{jhc84}.

For about 80\% of this sample we obtained an estimate of the physical
parameters (\teff, \logg) of stars and their related statistical uncertainty
by means of an original fitting procedure that matched the observed SED with
the ATLAS and NextGen model grids.

This provided a twofold application of our results; from one hand, we achieved
a self-consistent and independent calibration of the temperature and
bolometric scale  for giant and dwarf stars vs.\ empirical classification
parameters (i.e. spectral type and MK luminosity class). On the other hand,
the comparison of the synthetic templates from the ATLAS and NextGen model
grids allowed us to directly assess the relative performances of each
theoretical code to reproduce SED of real stars according to the different
input physics adopted.

The comparison of our results with several empirical calibration scales in 
the literature (see Sec.\ 4) led to the following main conclusions:

{\it i)} the good fitting accuracy ($\sigma_{\rm flux} \sim 2-5\%$) of both
theoretical models in reproducing SED of early-type stars (spectral type F and
earlier) drastically degrades at lower \teff, especially for K stars, where
both ATLAS and NextGen codes still fail to properly
account for the increasing contribution of molecular features in the spectra
of stars. In general, ATLAS is found to provide a systematically better fit (a
factor of two lower residual $\sigma_{\rm flux}$) than NextGen along the whole
B~$\to$~K spectral-type sequence, although the NextGen grid, due to its lower
\teff\ limit, more efficiently matches M stars.

{\it ii)} Comparing with empirical calibrations, both ATLAS and NextGen tend,
in average, to predict warmer (by 4--8\%) \teff\ for both giant and dwarf
stars of fixed spectral type. As for the ATLAS models, this effect has
probably much to do with the imperfect treatment of metal blanketing at short
wavelength \citep[as extensively discussed, for instance,
by][]{castellietal97}, while the case of NextGen seems more entangled.

This issue has been further explored in Sec.\ 5 by comparing the ATLAS vs.\
NextGen template sequences for 216 stars in the GS83 and JHC84 catalogs with
nominal best fit in the $3500 \leq \mteff \leq 10\,000$~K temperature
range. As a general feature, NextGen best-fit solutions are found to predict a
temperature and gravity excess with respect to the corresponding ATLAS
solutions for given target stars. The effect is especially evident for MK
I--III objects, where the NextGen fails to correctly settle \logg\ and
sensibly overestimates surface gravity in about 25\% of the cases vs.\ 4\% of
ATLAS. This misclassification partly derives from a lower capability of
NextGen spherical models to reproduce the SED of giant stars, compared to the
fit with the plane-parallel geometry. In most cases the latter proved in fact
to be formally more accurate, leading however to a less physical combination
of the fundamental parameters of stars.

An in-depth analysis of the fit accuracy for SED of target stars shows that,
to some extent, the NextGen \teff\ and \logg\ excess is correlated, as a
consequence of a sort of ``degeneracy'' in the solution space
\citep{buzzonietal01}. The effect is likely magnified in our framework when
considering that, for low-gravity stars, ATLAS model atmospheres assume
standard plane-parallel layers while NextGen adopts a spherical-shell
geometry. Because of a more important contribution of external atmosphere
layers to the integrated emerging flux in the case of the NextGen output, for
fixed \teff\ and \logg\ this implicitly calls for a ``redder'' theoretical SED
and a reduced blanketing absorption of metal blends at short wavelength, as a
consequence of ``sharper'' spectral features. 

The possible overestimate of the limb-darkening effects, as a consequence of
the adopted $T(\tau)$ vertical structure of NextGen model atmospheres, seems
also a critical issue in this regard, as indicated by the recent observations
of the EROS BLG2000-5 microlensing event.

\acknowledgments
It is a pleasure to thank the anonymous referee for his/her competent
suggestions, that greatly helped refining some important issues of our
discussion.

This work received partial financial support from the Italian MURST under grant
COFIN00 02-016, and from the Mexican CONACyT via grant 36547-E. This research
has made use of the SIMBAD database, operated at CDS, Strasbourg, France.


\end{document}